\documentclass[doublespacing]{elsart}
\journal{NIMA}
\usepackage{amssymb}
\usepackage{amsmath,amsfonts,verbatim,graphicx,float,xspace}
\usepackage{subfigure}

\usepackage{mdwlist}
\usepackage{lscape}
\usepackage{color}
\usepackage{wrapfig}
\usepackage{xspace}

\def\ruo{\rule{0mm}{4mm}}
\def\ruu{\rule[-2mm]{0mm}{6mm}}

\usepackage{cite}

\newfont{\tensy}{cmsy10}

\begin{document}
 
\begin{frontmatter}



\title{FADC Signal Reconstruction for the MAGIC Telescope}

 \author[a]{J.~Albert}, 
 \author[b]{E.~Aliu}, 
 \author[c]{H.~Anderhub}, 
 \author[d]{P.~Antoranz}, 
 \author[b]{A.~Armada}, 
 \author[d]{M.~Asensio}, 
 \author[e]{C.~Baixeras}, 
 \author[d]{J.~A.~Barrio},
 \author[f]{H.~Bartko\corauthref{cor1}}, 
 \ead{hbartko@mppmu.mpg.de}
 \author[g]{D.~Bastieri}, 
% ...

 \address[a]{Universit\"at W\"urzburg, D-97074 W\"urzburg, Germany}
 \address[b]{Institut de F\'\i sica d'Altes Energies, Edifici Cn., E-08193 Bellaterra (Barcelona), Spain}
 \address[c]{ETH Zurich, CH-8093 Switzerland}
 \address[d]{Universidad Complutense, E-28040 Madrid, Spain}
 \address[e]{Universitat Aut\`onoma de Barcelona, E-08193 Bellaterra, Spain}
 \address[f]{Max-Planck-Institut f\"ur Physik, D-80805 M\"unchen, Germany}
 \address[g]{Universit\`a di Padova and INFN, I-35131 Padova, Italy} 
% ...
 \corauth[cor1]{Corresponding author.}

% old author list

%% \thanks[label1]{}
%\author[munich]{H. Bartko\corauthref{cor1}},
%\corauth[cor1]{Corresponding author.}
%\ead{hbartko@mppmu.mpg.de}
%\author[padova]{M. Gaug\corauthref{cor1}},
%\ead{gaug@pd.infn.it}
%\author[munich]{F. Goebel},
%\author[IFAE]{A. Moralejo},
%\author[IFAE]{J. Rico},
%%\author[berlin]{M. Shayduk},
%\author[berlin]{Th. Schweizer},
%\author[IFAE]{N. Sidro},
%\author[munich]{W. Wittek}

%\address[munich]{Max Planck Institute for Physics, F{\"o}hringer Ring 6, 80805 Munich, Germany}
%\address[padova]{University of Padova, INFN, Via Marzolo 8, 35139 Padova PD, Italy}
%\address[IFAE]{IFAE Barcelona, Ed Cn, Campus UAB, 08193 Bellaterra, Spain}
%\address[berlin]{Humboldt University Berlin, Newtonstr. 15, 12489 Berlin, Germany}

%% abstract %%%%%%%%%%%%%%%%%%%%%%%%%%%%%%%%%%%%%%%%%%%%%%%%%
\begin{abstract}

Until April 2007 the MAGIC telescope used a 300~MSamples/s FADC system to sample the shaped PMT signals produced by the captured Cherenkov photons of air showers. Different algorithms to reconstruct the signal from the read-out samples (extractors) have been implemented and are described and compared. Criteria based on the obtained charge and time resolution/bias are defined and used to judge the different extractors, by applying them to calibration, cosmic and pedestal signals. The achievable charge and time resolution have been derived as functions of the incident number of photo-electrons.
%\begin{equation}
%\Delta T_{\mathrm{cosmic}} \approx \sqrt{\frac{(2\,\mathrm{ns})^2}{<Q>/{\mathrm{phe}}} 
%+ \frac{(4.5\,\mathrm{ns})^2}{<Q>^2/{\mathrm{phe^2}}} + (0.2\,\mathrm{ns})^2} . \nonumber
%\label{eq:time:fitprediction}
%\end{equation}
%For galactic backgrounds an image cleaning threshold as low as 5~photo-electrons can be achieved 
%without using the timing information and for rejecting 99.7\% of noise.

\end{abstract}

\begin{keyword}
fast digitization \sep FADC \sep digital filter \sep 
Cherenkov imaging telescopes, $\gamma$-ray astronomy.

\end{keyword}

\end{frontmatter}

% main text

\section{Introduction} 

The Major Atmospheric Gamma ray Imaging Cherenkov (MAGIC) telescope 
\cite{MAGIC-commissioning} uses the IACT technique~\cite{gamma_astro} %,low_energy}
 to study the very high energy (VHE, $E>50$~GeV) $\gamma$-ray emission from astrophysical sources, 
at the lowest possible energy threshold. The technique uses Cherenkov radiation:
A VHE $\gamma$-ray entering the earth's atmosphere initiates a shower (cascade) of electrons and positrons, 
with a particle density maximum about 10~km above sea level (for an energy of 1~TeV). 
The particles in the cascade produce Cherenkov light in a cone of about $1^{\circ}$ 
half-angle, which illuminates an area of around 120~m radius
on the ground. If the MAGIC telescope is located in this area, 
part of the Cherenkov light 
will be collected by the telescope mirrors and a shower image 
will be projected onto the photomultiplier tube (PMT) camera.
The Cherenkov photons arrive within a very short time interval of a few nanoseconds at the telescope camera, whose pixels are fast light sensors such as PMTs, so that 
one can trigger on the coincident light signals. The fluctuations of the light of the night sky (LONS) cause background noise. This effect is minimized by using low exposure times (signal integration times), typically of the order of ten nanoseconds.

To reach the highest sensitivity and the lowest energy threshold, the recorded signals have to be accurately reconstructed. Two quantities are of interest: The  total signal charge and the signal arrival time. The signal charge (the total number of photo-electrons released from the photocathode of the PMT) is proportional to the total area below the pulse. The sum of the signal charges of all camera pixels is a measure of the shower energy. The signal arrival time
 is given by the time difference between the first recorded FADC sample and a characteristic position on the pulse shape, like the maximum, the half-maximum on the rising edge or the center of gravity of the pulse.
%The highest possible signal-to-noise ratio and signal charge/time resolution, as well as a small bias are important.

The timing information may be used to discriminate between pixels whose signals belong 
to the shower, and pixels which are affected by 
randomly timed background noise. The pixels with a low signal-to-noise ratio are 
rejected for the subsequent image parameterization \cite{Hillas_parameters,Fegan1997}.

The main background to $\gamma$-rays originates from the much more frequent showers induced by isotropic 
hadronic cosmic rays.
Monte Carlo (MC) based simulations predict different time structures for $\gamma$-ray and hadron 
induced shower images as well as for images of single muons 
\cite{muon_rejection,MC_timing_Indians,Roberts1998,Hess1999}. This has two consequences: 
On the one hand the arrival time structures across the observed Cherenkov shower image, 
from pixel to pixel, depend on the type of the primary particle. 
On the other hand, also the recorded Cherenkov pulse shape inside an individual pixel depends 
on the primary particle. To exploit the pulse shape differences, an ultra-fast digitization of the 
Cherenkov pulses is necessary, as is provided by the most recent upgrade of the data acquisition of the 
MAGIC telescope to a 2~GSamples/s FADC system \cite{GSamlesFADC,MuxInstallation}. This paper, however, deals with the signal 
reconstruction of the data taken with the initial 
300~MSamples/s FADC system. Because of its limited sampling speed, we do not try to exploit the 
differences in pulse shape here.

%An accurate arrival time and Cherenkov pulse shape determination may therefore improve the $\gamma$/hadron separation power \cite{MC_timing_Indians,Roberts1998,Hess1999}. According to \cite{Hess1999} the pulse shape differences can only 

%Moreover, the timing information may be used to discriminate between pixels whose signal belongs to the shower and pixels which are affected by randomly timed background noise. The pixels with a low signal-to-noise ratio are rejected for the subsequent image parameterization \cite{Hillas_parameters,Fegan1997}.

This paper is structured as follows: In section~\ref{sec:signal_read_out} the read-out system of the 
MAGIC telescope is described, and in section~\ref{sec:reco} the average pulse shapes of calibration and 
cosmic pulses are reconstructed, from data taken with the FADC system. 
These pulse shapes are compared to those implemented in the MC simulation program. 
In section~\ref{sec:criteria} criteria for an optimal signal reconstruction are developed.
In section~\ref{sec:algorithms} the signal reconstruction algorithms 
and their implementation in the MAGIC software framework ({\textit{\bf MARS}}~\cite{MARS}) are described. 
The performance of the signal extraction algorithms under study is assessed by applying them to pedestal, 
calibration and MC events (sections~\ref{sec:mc} to~\ref{sec:calibration}). 
Section \ref{sec:speed} gives the CPU time requirements for the different signal reconstruction algorithms. 
Finally in sections~\ref{sec:Extraction:results} and \ref{sec:outlook} the results are summarized
and an outlook is given.

\section{Signal read-out}\label{sec:signal_read_out}

\begin{figure}[h!]
\begin{center}
\includegraphics[width=\textwidth]{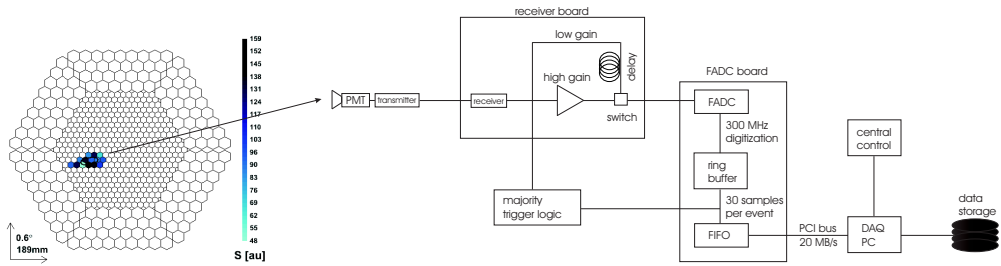}
\end{center}
\caption[MAGIC read-out scheme.]{MAGIC read-out scheme: the analog PMT signals are 
transferred via an analog optical link to the counting house where -- after the trigger decision -- the signals 
are digitized by a 300\,MHz FADCs system and written to the hard disk of a data acquisition PC.} 
\label{fig:MAGIC_read-out_scheme}
\end{figure}

Figure~\ref{fig:MAGIC_read-out_scheme} shows a sketch of the MAGIC read-out system, including the PMT camera, the analog-optical link, the majority trigger logic and flash analog-to-digital converters (FADCs). 
%The used PMTs (ET\,9116A from Electron Tubes) provide a very fast response to the input light signal. 
The response of the PMTs to sub-ns input light pulses shows a full width at half maximum (FWHM) of 1.0 - 1.2~ns and rise and fall times of 600 and 700\,ps correspondingly~\cite{Magic-PMT}. 
A transmitter using a vertical-cavity surface-emitting laser (VCSEL) diode modulated in amplitude, 
converts the electrical pulse supplied by the PMT into an optical signal. 
This signal is then transferred via optical fibers (162\,m long, 50/125\,$\mu$m diameter) 
to the counting house~\cite{MAGIC-analog-link-2}. After transforming the light back to an electrical signal, 
the original PMT pulse has a FWHM of about 2.2~ns for a single photo-electron pulse, 
and rise and fall times of about 1\,ns. % was 2.2 ns

%an analog optical link \cite{MAGIC-analog-link-2} to the counting house.

%After modulating VCSEL type laser diodes, after traveling through 162m of multi-mode graded index fiber of 50/125 $\mu$m diameter and.

%In order to match the bandwidth of the DAQ system and to suppress frequencies above the Nyquist frequency of the digitizers the fast PMT signals are shaped

\begin{figure}[htp]
\begin{center}
\includegraphics[width=0.485\linewidth]{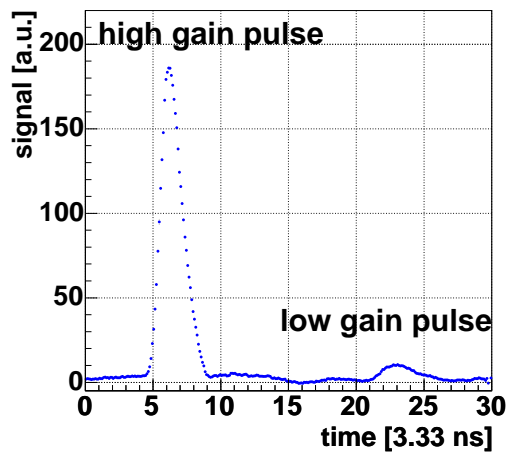}%{pulpo_shape_high.eps}
\end{center}
\caption[Reconstructed high gain shape.]{Average reconstructed pulse shape from a fast pulse generator showing the high gain and the low gain pulse. The FWHM of the high gain pulse is about 6.3\,ns while the FWHM of the low gain pulse is about 10\,ns.} 
\label{fig:pulpo_shape_high}
\end{figure}

In order to sample this pulse shape with the 300 MSamples/s FADC system, the original pulse is electronically 
shaped by effectively folding it with a 
%with a shaping 
function of 6\,ns FWHM. In order to increase the dynamic range of the read-out, the signals 
are split into two branches, with gains differing by a factor~10. The low-gain branch is delayed by 55\,ns 
and both branches are multiplexed and read out by one FADC. The switch from high- to low-gain occurs only
if the high-gain signal exceeds a pre-set threshold, and 
55\,ns after this happens. During the subsequent 50\,ns the low-gain signal is connected to the
output while the high-gain signal is blocked.
% the high-gain channel is blocked. 
Figure~\ref{fig:pulpo_shape_high} shows the average reconstructed pulse shape (generated by a fast pulser, 
see section \ref{sec:reco}) as measured by one FADC. A more detailed overview about the MAGIC read-out 
and DAQ system can be found in \cite{Magic-DAQ}. 
%Since April 2007  Planned upgrades of the read-out system to a higher sampling speed are described in \cite{GSamlesFADC,domino}. 
% The maximum sustained trigger rate could be 1 kHz. The FADCs feature a FIFO memory which allows a significantly higher short-time rate.
% Obviously by doing this, more LONS is integrated and thus the performance of the telescope on the analysis level is degraded. 

% \subsection{Characteristics of the current read-out system}

The following intrinsic characteristics of the MAGIC read-out system are the most important to
affect the signal reconstruction:

%\begin{description}
%\begin{itemize}
%\item[Inner and Outer pixels:\xspace] 
{\bf Inner and Outer pixels: }
The MAGIC camera is constructed with two types of pixels, inner and outer pixels, 
%(see Figure \ref{fig:MAGIC_read-out_scheme}),
 with the following differences:
\begin{enumerate}
\item{Size: The outer pixels have an area larger than the inner pixels by a factor four~\cite{MAGIC-design}. 
Their area multiplied by photon detection efficiency, however, is higher only by a factor~2.6.%quantum-efficiency
%Their (quantum-efficiency convoluted) effective area is about a factor 2.6 higher.
}
\item{Gain: The camera is flat-fielded in order to yield a similar reconstructed charge signal in all pixels, for the same photon illumination intensity. 
In order to achieve this, the gain of the inner pixels has been adjusted to about a factor 2.6 higher than the outer 
ones~\cite{MAGIC_calibration}. This results in a lower charge RMS contribution from the light of the night sky (LONS) for the outer pixels.}
% effective noise charge from the night sky background for the outer pixels.} 
\item{Delay: Due to the lower high voltage (HV) settings of the outer pixels, their signals are delayed by about 1.5\,ns 
with respect to the inner ones.}
\end{enumerate}

%\item[Asynchronous trigger:\xspace] 
{\bf Asynchronous trigger: } 
The FADC clock is not synchronized with the trigger. Therefore the time $t_{\mathrm{rel}}$ between the trigger decision and the first read-out sample is uniformly distributed along the range $t_{\mathrm{rel}} \in [0,T_{\mathrm{FADC}}]$, where $T_{\mathrm{FADC}}=3.33$\,ns is the digitization period of the MAGIC 300\,MHz FADCs.
% answer to the referee:
All FADCs run at the same frequency and phase. The 300 MHz clock signal is produced at a central place, multiplicated and distributed by equally long cables to the individual FADC modules.

%\item[AC coupling:\xspace] 
{\bf AC coupling: }
The PMT signals are AC-coupled at various places in the signal transmission chain. Thus the contribution of the PMT pulses due to the LONS is on average zero. Only the signal RMS depends on the intensity of the LONS. In moonless nights, observing an extra-galactic source, an average background rate of about 0.13 photo-electrons per nano-second per inner pixel has been measured~\cite{GSamlesFADC}.
%\begin{figure}[h!]
%\begin{center}
%\includegraphics[totalheight=7cm]{shape_25945_raw.eps}
%\end{center}
%\caption[Raw shape.]{Superimposed FADC slices of 1000 pulse generator pulses.} 
%\label{fig:raw_shape}
%\end{figure}

%\begin{figure}[h!]
%\begin{center}
%\includegraphics[totalheight=7cm]{time_25945.eps}
%\end{center}
%\caption[Reconstructed time.]{Distribution of the reconstructed arrival time from the raw FADC samples shown in Figure \ref{fig:raw_shape}. The distribution has a width of about 1 FADC period (3.33\,ns) due to the asynchronous trigger with respect to the FADC clock.
%The width of the distribution is due to the trigger jitter of 1 FADC period (3.33 ns).
%} 
%\label{fig:reco_time}
%\end{figure}

%Figure~\ref{fig:raw_shape} shows the raw FADC values as a function of the slice number for 1000 constant pulse generator pulses overlayed and figure~\ref{fig:reco_time} shows the distribution of the corresponding reconstructed arrival times. The distribution has FWHM of about 1 FADC period (3.33\,ns) due to the asynchronous trigger with respect to the FADC clock.

%\item[Shaping:\xspace] 
{\bf Shaping:}
  As already mentioned above, the optical receiver board shapes the pulse with a shaping time of 6\,ns FWHM, i.e. much larger than the typical intrinsic pulse width.
%  The optical receiver board shapes the pulse with shaping times of about 6\:ns, much larger than the typical intrinsic pulse width. 
Since the shaping time is larger than the width of a single FADC slice, a strong correlation of the noise between neighboring FADC slices is expected.

%%\item[Instantaneous amplitudes:\xspace] 
%{\bf Instantaneous amplitudes: }
%  The MAGIC FADCs consist of a series of small comparators which measure the instantaneous amplitude 
%of a pulse at a time. No charge integration over the duration of a time slice is performed by the FADCs. 
%Therefore, pulse structures with a frequency higher than 150\,MHz are lost.
%%\end{description}
%%\end{itemize}

\section{Pulse Shape \label{sec:reco}}

% The relative position of the recorded signal samples varies from event to event with respect to the position of the signal shape. The time 

%As the FADC clock is not synchronized with the trigger, the time $t_{\mathrm{rel}}$ between the trigger decision and the first read-out sample is uniformly distributed in the range $t_{\mathrm{rel}} \in [0,T_{\mathrm{FADC}}]$, where $T_{\mathrm{FADC}}=3.33$\,ns is the digitization period of the MAGIC 300\,MHz FADCs. $t_{\mathrm{rel}}$ can be determined using the reconstructed arrival time $t_{\mathrm{arrival}}$.
%directly by a time to digital converter (TDC) or 

The fact that the signal pulses are sampled asynchronously by the FADCs allows one to determine the average pulse shape with high accuracy. To do that, the signal samples from different recorded pulses are shifted to a common reconstructed arrival time and normalized to a common area/charge. Therefore, the precision of the determination of each point along the pulse shape depends on the accuracy of the arrival time and charge reconstruction. Possible biases in the charge and arrival time reconstruction may introduce systematic errors, whose size are unknown at first hand.
% It is used to simulate the response of the photo-multipliers to Cherenkov light. 
Figure~\ref{fig:pulpo_shape_high} shows the average signal from a fast pulser as reconstructed 
by the MAGIC read-out system. The relative statistical error of the amplitude value of every reconstructed 
point is well below $10^{-2}$. The pulser generates unipolar pulses of about 2.5~ns FWHM and 
with a preset amplitude. These electrical pulses are fed to the VCSEL transmitters, and are transmitted 
using the same analog-optical link as the PMT pulses, and are fed into the MAGIC receiver board. 
%The pulse generator setup is mainly used for test purposes of the receiver board, trigger logic and FADCs. %In figure~\ref{fig:pulpo_shape_high} the high and the low gain pulses are clearly visible. The low gain pulse is attenuated by a factor of $\sim 10$ and delayed by  $\sim 55$\,ns with respect to the high gain pulse.

\begin{figure}[h!]
\begin{center}
\includegraphics[width=0.485\linewidth]{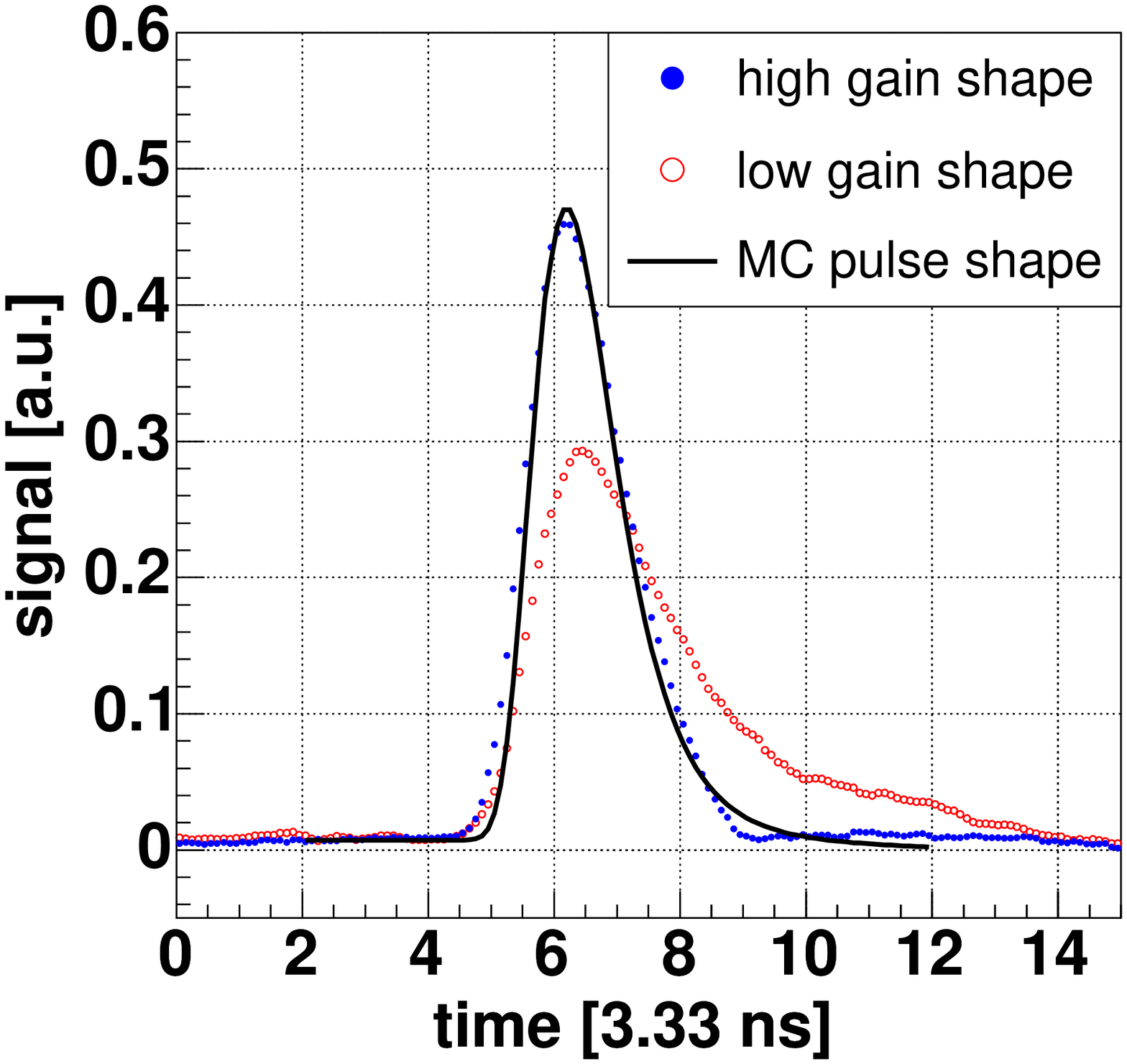}%{pulpo_shape_low.eps}
\includegraphics[width=0.485\linewidth]{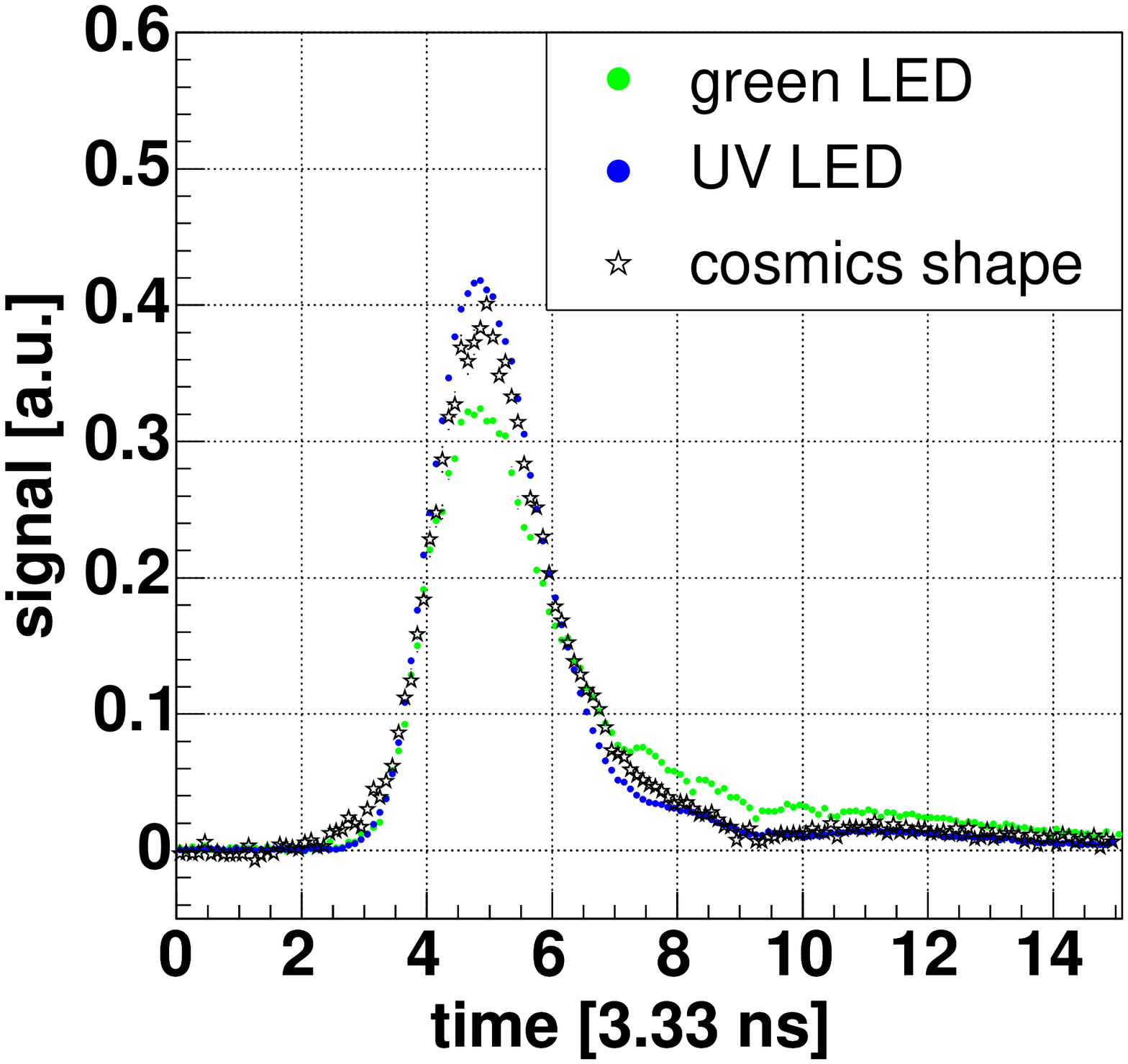}%{shape_green_high.eps} % was totalheight=7cm
\end{center}
\caption[Reconstructed pulse shapes]{%
\label{fig:pulse_shapes}
Left: Average reconstructed high gain and low gain pulse shapes from a pulse generator run. The black line corresponds to the pulse shape implemented into the MC simulations \cite{Majumdar2005}.
Right: Average reconstructed high gain pulse shape for calibration runs with green and UV light (see section~\ref{sec:calibration}). All pulse shapes are normalized to a common arrival time and area.} 
\end{figure}

Figure~\ref{fig:pulse_shapes} (left) shows the normalized average pulse shape for the pulse generator in the high and in the low gain, respectively. The intrinsic FWHM of the generated pulses is 2.5\,ns, whereas it is on average 6.3\,ns and 10\,ns for the pulses reconstructed from the high and low gain chains, respectively. The broadening of the low gain pulses with respect to that of the high gain is due to the limited bandwidth of the passive 55\,ns on-board delay line of the MAGIC receiver boards. 
%   while the FWHM of the average reconstructed low gain pulse shape is 
% Due to the electric delay line for the low gain pulses on the receiver board the low gain pulse is widened with respect to the high gain. 
%The low gain pulse has a FWHM of about 10\,ns.

Figure~\ref{fig:pulse_shapes} (right) shows the normalized average reconstructed pulse shapes for green and UV calibration LED pulses~\cite{MAGIC-calibration}
 (see section \ref{sec:calibration}) as well as that of cosmic events. The shapes of the UV calibration and cosmic pulses are quite similar. Both have a FWHM of about 6.3 ns. Since air showers from hadronic cosmic rays trigger the telescope much more frequently than $\gamma$-ray showers, the reconstructed pulse shape of the cosmic events corresponds mainly to hadron induced showers. The pulse shape from electromagnetic air showers might be slightly different as indicated by MC simulations \cite{MC_timing_Indians,muon_rejection}. The pulse shape for green calibration LED pulses is wider and has a pronounced tail. The difference between the shapes of the calibration LED pulses is not due to the LED light color but to different electronics used for the fast LED drivers.

%Having reconstructed the pulse shapes from generator pulses, cosmic and calibration pulses
The reconstructed pulse shapes for generator pulses, cosmic and calibration events permit to implement 
a representative pulse shape in the MC simulations, see e.g. the full black line in 
figure \ref{fig:pulse_shapes}, left panel. 
%Also the difference between the FWHM of the pulses in the high and the low gain has to be taken into account for the MC simulations and for the signal extraction algorithms. Moreover t
The shape difference between the calibration pulses and the cosmic pulses has to be corrected for in the calibration procedure \cite{MAGIC_calibration}.

\section{Criteria for Optimal Signal Extraction \label{sec:criteria}}

The goal of the optimal signal reconstruction algorithm is to compute an unbiased estimate of the charge and arrival time of the Cherenkov pulse with the highest possible resolution. 
%An accurate determination of the arrival time may help distinguish between signal and background. The signal arrival time varies smoothly from pixel to pixel while the background noise is randomly distributed in time. 
%Therefore, it must be insured that the reconstructed arrival time corresponds to the same reconstructed pulse as the reconstructed charge.
% From Markus
%In the following, a brief introduction to the theory of signal extraction is given, 
%
Let us consider a large number of identical signals, corresponding to a fixed number of photo-electrons $N_{\mathrm{phe}}$. By applying a signal extraction algorithm, a distribution of estimated signals $\widehat{N_{\mathrm{phe}}}$ is obtained, see also \cite{Smith1997} and references therein. Criteria for an optimal signal reconstruction algorithm are developed according to~\cite{james}.
% (for fixed $N_{\mathrm{phe}}$ and fixed background fluctuations {\textit{BG}). 
The deviation between true and reconstructed value is given by:
\begin{equation}
X = \widehat{N_{\mathrm{phe}}}- N_{\mathrm{phe}} \ .
\end{equation}
The distribution of $X$ has a mean $B$ (the {\bf Bias} of the estimator) and a variance $V$.% defined as:
%
%\begin{eqnarray}
%   B   \ \ \ \  \equiv \ \ \ \ \ \ <X> \ \ \ \ \  &=& \ \ <\widehat{N_{\mathrm{phe}}}> \ -\ N_{\mathrm{phe}}\\
%   V \ \ \ \ \equiv \ <(X-B)^2> &=& \ \mathrm{Var}[\widehat{N_{\mathrm{phe}}}] \label{eq:def:r}\\
%\textit{RMSE}\ \ \equiv \ \ \sqrt{<X^2>} \ \ \ \  &=& \ \sqrt{V+B^2} \ .
%\end{eqnarray}
%
The parameter $B$ is also called the {\bf Bias} of the estimator and 
\textit{RMSE}  is the {\bf Root Mean-Squared Error} which combines resolution and bias:
\begin{equation}
\textit{RMSE}\ \ \equiv \ \ \sqrt{<X^2>} \ \ \ \  = \ \sqrt{V+B^2} \ .
\end{equation}
%Typically, one measures easily the parameter $V$, but needs the \textit{RMSE} for statistical analysis. However, only in case of a vanishing bias, the two numbers are equal.
%Otherwise, the bias has to be determined. 

Generally, both $B$ and \textit{RMSE} depend on $N_{\mathrm{phe}}$ and the background fluctuations \textit{BG}. 
In the case of the MAGIC telescope, the background fluctuations are due to the electronics noise and the PMT response to the LONS. The signals from the latter have the same shape as those from Cherenkov pulses. Therefore, those algorithms which search for the highest sum of a number of consecutive FADC slices inside a global time window (so-called {\bf sliding window algorithms}) will have a bias. In case of no Cherenkov signal they will typically reconstruct the largest noise pulse. Nevertheless, such a sliding window algorithm usually has a much smaller variance and in many cases a smaller \textit{RMSE} than the {\bf fixed window extractors}, which just sum up a fixed number of FADC slices. 
%For the description of the signal reconstruction algorithms see section \ref{sec:algorithms}.

%Note that every signal extraction algorithm which searches for signals inside a global time window (a so-called {\bf sliding window algorithm}) will hence have a bias, especially at low or vanishing signals, but usually a much smaller variance and in many cases a smaller \textit{RMSE} than the {\bf fixed window extractors}.

%\subsection{Linearity}

The reconstructed charge should be proportional to the total number of photo-electrons in the PMT. 
This linearity is very important for the reconstruction of the shower energy and hence for the measurement 
of energy spectra from astronomical sources. Deviations from linearity may be caused in different ways:  
At very low signals, the signal will be biased towards too high values (positive $X$). 
At very high signals, the FADC system goes into saturation, and the reconstructed signal becomes 
too low (negative $X$). Also, any error in the inter-calibration between the high and low gain acquisition 
channels yields an effective deviation from linearity.

%\begin{itemize}
%\item At very low signals, the bias causes too high a signal to be reconstructed (positive $X$).
%\item At very high signals, the FADC system goes into saturation and the reconstructed signal becomes too low (negative $X$).
%\item Any error in the inter-calibration between the high and low gain acquisition channels yields an effective deviation from 
%linearity.
%\end{itemize}

%\subsection{Robustness}

Another important feature of an extractor is its robustness, i.e. its stability in reconstructing the charge and arrival time for
% different pulse shapes. The signal extraction algorithms have to reconstruct stably the charge and arrival time for 
different types of pulses with different intrinsic shapes and background levels: 

\begin{itemize}
\item{Cherenkov signals from $\gamma$-rays, hadrons and muons}
\item{calibration pulses from different LED color pulsers (with different pulse shapes, see figure \ref{fig:pulse_shapes} right panel)}
\item{pulse generator pulses.}
\end{itemize}

%\subsection{Low Gain Extraction}
Finally, the extractor has to accurately reconstruct both the high and low gain channels. 
Due to the analog delay line, the low gain pulse is wider. % and the charge spans a longer time window. 
%The time delay between the tail of the high gain pulse and the rising edge of the low gain pulse is small. Thus for large pulses, mis-interpretations between the tail of the high gain pulse and the low gain pulse might occur. Moreover, t
The total recorded time window is relatively small, such that parts of the low gain pulse may 
lie outside of the recorded FADC window. 
%A good extractor must stably extract the low gain pulse without being confused by the above issues. 

%J. Rico: remove
%An important point is the difference between the pulse shapes of the calibration and Cherenkov signals. It has to be ensured that the computed calibration factor between the reconstructed charge in FADC counts and photo-electrons for calibration events is also valid for signals from Cherenkov photons.

%\subsection{Applicability for Different Sampling Speeds / No Pulse Shaping.}

%The current read-out system of the MAGIC telescope~\cite{Magic-DAQ} with 300~MSamples/s is relatively slow compared to the fast Cherenkov pulses of about 2\,ns FWHM. An artificial pulse shaping to about 6.5\,ns FWHM is used and thereby also more LONS is integrated. 

%For 2 ns FWHM fast pulses a 2 GSamples/s FADC provides at least 4 sampling points. This permits a reasonable reconstruction of the pulse shape. First prototype tests with fast digitization systems for MAGIC have been successfully conducted~\cite{GSamlesFADC,domino}. 

\section{Signal Reconstruction Algorithms}\label{sec:algorithms}

We have chosen four algorithms for the study of the reconstruction of the signal charge and arrival time: Fixed Window, Sliding Window with Amplitude-weighted Time, Cubic Spline with Integral or Amplitude Extraction, and Digital Filter. For the signal reconstruction algorithms adopted by other air Cherenkov telescopes, see e.g. \cite{Holder2005,Holder2006,Hess1999,Cogan2007}.

%\begin{itemize}
%\item{Fixed Window}
%\item{Sliding Window with Amplitude-weighted Time}
%\item{Cubic Spline with Integral or Amplitude Extraction}
%\item{Digital Filter.}
%\end{itemize}

\subsection{Fixed Window}

This signal extraction algorithm simply adds the pedestal-subtracted FADC slice contents of a fixed range (window) of consecutive FADC slices. The window has to be chosen large enough to always cover the complete pulse, otherwise physical differences in the pulse position with respect to the FADC slice numbering would lead to integration of different parts of the pulse. For this reason, the fixed window algorithm adds up more noise than the other considered signal reconstruction algorithms. Due to the AC-coupling of the read-out chain, the reconstructed signals have no bias.

In the current implementation, the fixed window algorithm does not calculate arrival times.

\subsection{Sliding Window with Amplitude-weighted Time}

This signal extraction algorithm searches for the maximum integral content among all possible FADC windows of fixed size contained in a defined time range (global window). The arrival time is calculated from the window with the highest integral as:
\begin{equation} \label{eq:sliding_window_time}
  t = \frac{\sum_{i=i_0}^{i_0+\mathrm{\it ws}-1} s_i \cdot t_i}{\sum_{i=i_0}^{i_0+\mathrm{\it ws}-1} s_i} \ ,
\end{equation}
where $i$ denotes the FADC slice index, starting from slice $i_0$ and running over a window of size $\mathrm{\it ws}$. The $s_i$ are the pedestal-subtracted FADC slice contents and the $t_i$ are the corresponding times relative to the first recorded FADC slice. 

\subsection{Cubic Spline with Integral or Amplitude Extraction}

This signal extraction algorithm interpolates all the pedestal-subtracted FADC slice contents of the full read-out window
using a cubic spline algorithm, adapted from~\cite{NUMREC}. In a second step, it searches 
for the position of the maximum of the interpolation function. Thereafter, 
two different estimators of the pulse charge are available:

\begin{enumerate}
\item{{\bf Amplitude}: the value of the spline maximum is taken as reconstructed signal.}
\item{{\bf Integral}: The interpolation function is integrated in a window of fixed size, with integration limits 
fixed with respect to the position of the spline maximum.}
\end{enumerate}

The pulse arrival times can also be estimated in two ways:

\begin{enumerate}
\item{{\bf Pulse maximum}: The position of the spline maximum determines the arrival time.}
\item{{\bf Pulse Half Maximum}: The position of the half maximum at the rising edge of the pulse determines the arrival time.}
\end{enumerate}

\subsection{Digital Filter}

The goal of the digital filtering method~\cite{OF94,OF77} is to optimally reconstruct the charge and arrival time of a signal whose shape is known. Thereby, the noise contributions to the amplitude and arrival time reconstruction are minimized, see also~\cite{DF_astro-ph}. 
For the digital filtering method to work properly, two conditions have to be satisfied:

\begin{itemize}
\item{The normalized signal shape has to be constant.}
\item{The noise properties must be constant, i.e. the noise is stationary and independent of the signal amplitude.}
%\item{The normalized noise auto-correlation has to be constant, i.e. the noise is stationary.}
\end{itemize}

% This will be shown later explicitly.
As the pulse shape in MAGIC is mainly determined by the artificial shaping
% pulse stretching 
on the optical receiver board, the first assumption holds to a good approximation for 
all pulses with intrinsic signal widths much smaller than the shaping constant. 
Also the second assumption is satisfied to a good approximation: 
Signal and noise are independent and the measured pulse is a linear superposition of the 
signal and noise contributions. 
%The validity of the third assumption is discussed below, especially for different 
LONS conditions.

% (usually smaller than one FADC slice width),
Let $g(t)$ be the normalized signal shape (e.g. from figure~\ref{fig:pulse_shapes}), $E$ the signal integral (charge) and $\tau$ the shift between the timing of the physical and the considered/probed signals. Then the time dependence of the signal is given by $y(t)=E \cdot  g(t-\tau) + b(t)$, where $b(t)$ is the time-dependent noise contribution. For small time shifts $\tau$ the time dependence can be linearized. Discrete measurements $y_i$ of the signal at times $t_i \ (i=1,...,n)$ have the form $y_i=E \cdot g_i- E\tau \cdot \dot{g}_i + O(\tau^2) +b_i$, where $\dot{g}(t)$ is the time derivative of the signal shape, $g_i=g(t_i)$ and $b_i=b(t_i)$.
%
%\begin{equation}
%y_i=E \cdot g_i- E\tau \cdot \dot{g}_i +b_i \ .
%\end{equation}

%\begin{equation} \label{shape_taylor_approx}
%y(t)=E \cdot  g(t) - E\tau \cdot  \dot{g}(t) + b(t) \ ,
%\end{equation}

%\begin{figure}[htp]
%\begin{center}
%\includegraphics[totalheight=6.5cm]{noise_38995_smallNSB_all396.eps}
%\includegraphics[totalheight=6.5cm]{noise_39258_largeNSB_all396.eps}
%\includegraphics[totalheight=6.5cm]{noise_small_over_large.eps}
%\end{center}
%\caption[Noise autocorrelation average all pixels.]{a,b) Noise autocorrelation matrix $\boldsymbol{B}$ averaged over all pixels for two different LONS levels: a) telescope pointing off the galactic plane, b) telescope pointing into the galactic plane. c) The ratio between a) and b). One can see that the entries of $\boldsymbol{B}$ do not simply scale with the amount of night sky background.} 
%\label{fig:noise_autocorr_allpixels}
%\end{figure}

The correlation of the noise contributions at times $t_i$ and $t_j$ can be expressed by the noise autocorrelation matrix
\begin{equation}
\boldsymbol{B}: B_{ij} = \langle b_i b_j \rangle - \langle b_i \rangle \langle b_j\rangle \ ,  
\label{eq:Extraction:autocorr}
\end{equation}
whose elements can be obtained from pedestal data (see section \ref{sec:pedestals}). 
%Figure~\ref{fig:noise_autocorr_allpixels} shows the measured noise autocorrelation matrix for different LONS levels. 
The noise autocorrelation matrix is dominated by LONS pulses shaped by 6\,ns FWHM. The absolute scale of the matrix elements depends on the LONS level. The normalized matrix elements may change by about 10\% due to varying LONS levels in typical observation conditions.
The noise auto-correlation in the low gain channel cannot be determined from data. The low gain channel read-out is only activated in case the high gain signal is above a certain threshold resulting in a measurable low gain signal. It has to be retrieved from Monte-Carlo studies instead.

\begin{figure}[htp]
\centering
\includegraphics[width=0.85\linewidth]{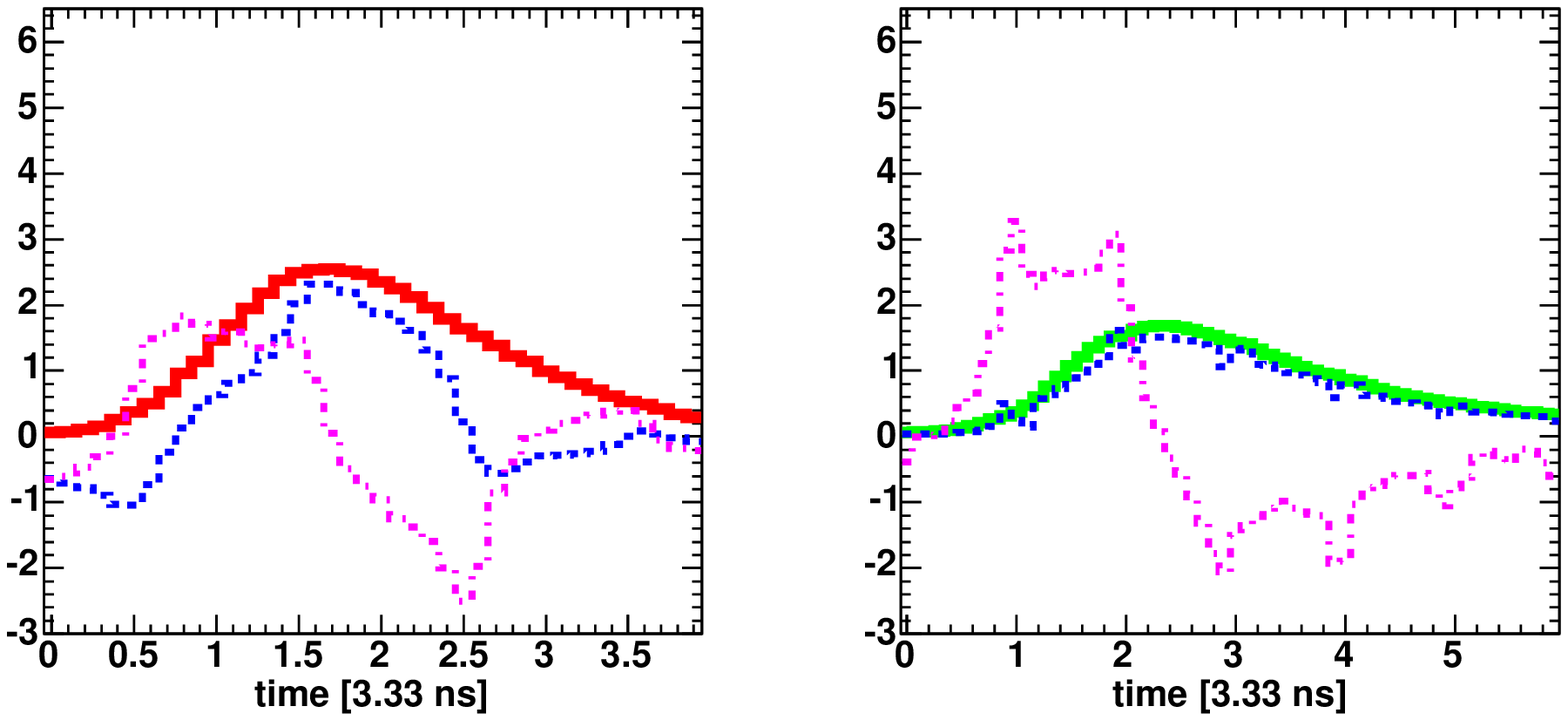}
%% \caption[Digital Filter Weights Pulses]{%
%% \label{fig:Extraction:weightscosmic}
%% Example of digital filter weights, used for cosmic pulses. 
%% Left: high gain, right: low gain. Full lines: normalized signal shapes $g(t)$ (multiplied 
%% with 5 for better visibility), dashed lines: amplitude weights $w_{\mathrm{amp}}(i_0)$, 
%% dotted-dashed: time weights $w_{\mathrm{time}}(i_0)$. }
\vspace{\floatsep}
\includegraphics[width=0.85\linewidth]{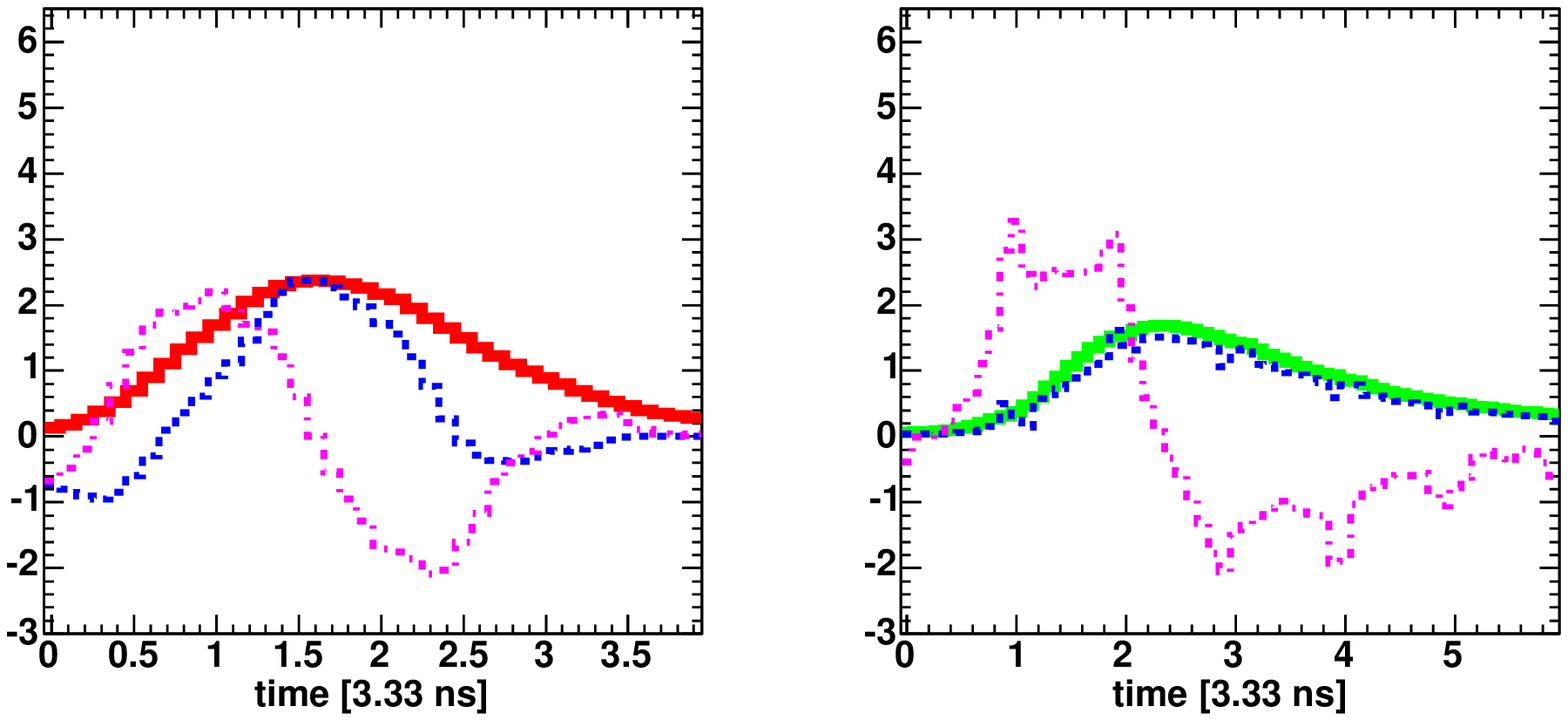}
%% \caption[Digital Filter Weights Calibration Pulses UV]{%
%% \label{fig:Extraction:weightscalibuv}
%% Example of digital filter weights, used for UV calibration pulses. 
%% Left: high gain, right: low gain. Full lines: normalized signal shapes $g(t)$ (multiplied 
%% with 5 for better visibility), dashed lines: amplitude weights $w_{\mathrm{amp}}(i_0)$, 
%% dotted-dashed: time weights $w_{\mathrm{time}}(i_0)$. }
\vspace{\floatsep}
\includegraphics[width=0.85\linewidth]{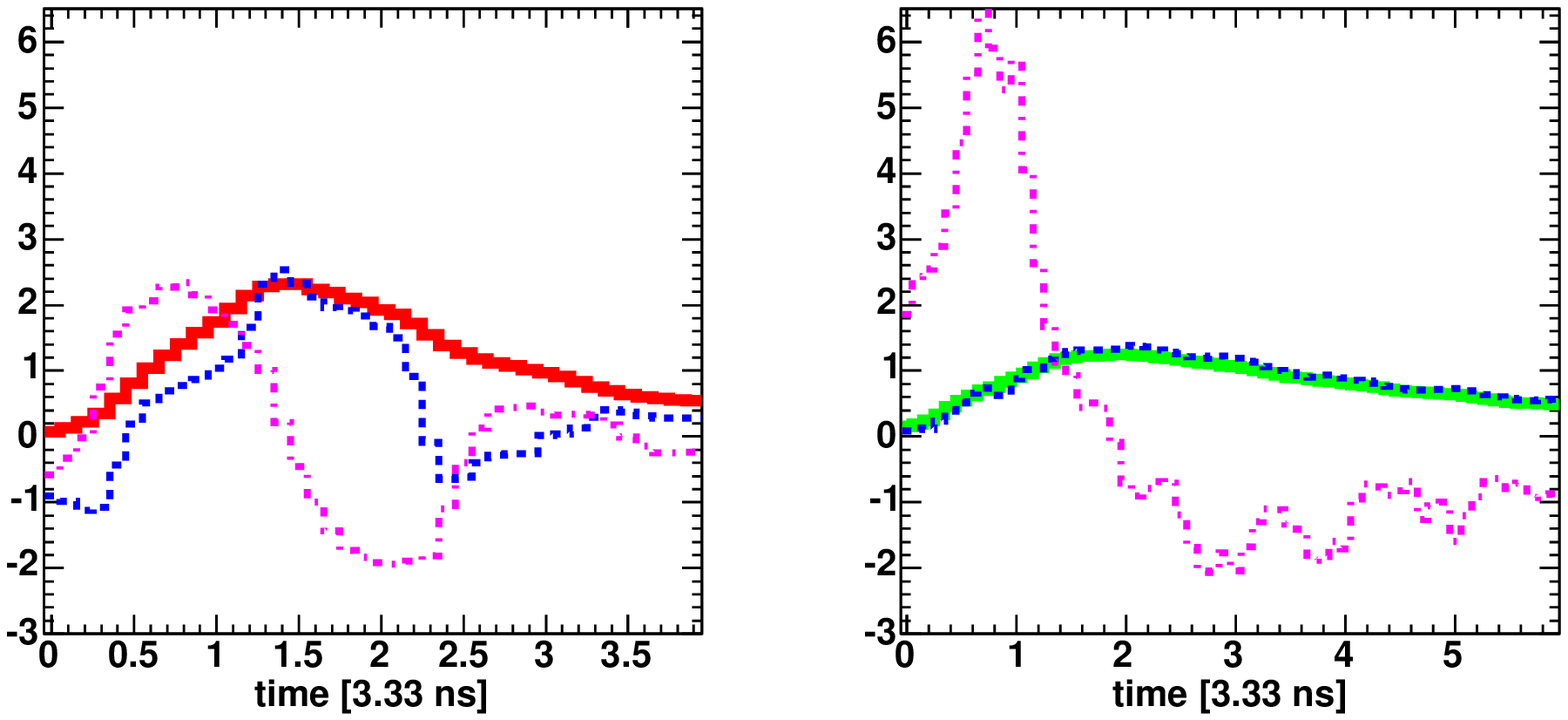}
\caption[Digital Filter Weights]{% Calibration Pulses Blue/Green]{%
\label{fig:Extraction:weights}
%\label{fig:Extraction:weightscalibblue}
Examples of digital filter weights. Top: cosmic pulses, center: UV calibration pulses and bottom: blue and green calibration pulses. On the left side, the high gain pulse is shown, one the right side, the low gain. Full lines show the normalized signal shapes $g(t)$ (multiplied by 5 for better visibility), dashed lines the amplitude weights $w_{\mathrm{amp}}(t)$, and dotted-dashed lines the time weights $w_{\mathrm{time}}(t)$. For the high-gain extraction 4 FADC slices are used and for the low-gain extraction 6 FADC slices.}
%\includegraphics[width=0.45\linewidth]{w_time_MC_input_TDAS.eps}
%\caption[Digital Filter time weights]{%
%\label{fig:Extraction:timeweights}
%Example of time weights $w_{\mathrm{time}}(\tau)|_{i_0}$ for a window size of 6 FADC slices. } 
%\end{figure}
%\begin{figure}[h!]
%\begin{center}
%\hspace{\floatsep}
%\caption[Digital Filter Weights Calibration Pulses]{%
%\label{fig:Extraction:weightscosmic}
%Example of digital filter weights  for a window size of 6 FADC slices. 
%Left: $w_{\mathrm{amp}}(\tau)|_{i_0}$, right: $w_{\mathrm{time}}(\tau)|_{i_0}$. } 
\end{figure}

For a given pulse, $E$ and $E \tau$ can be estimated from the $n$ FADC measurements $\boldsymbol{y} = (y_1, ... ,y_n)$ by minimizing the deviation between the measured and the known pulse shape, and taking into account the known noise auto-correlation, i.e. minimizing the following expression (in matrix form): 
\begin{equation}
\chi^2(E, E\tau) = (\boldsymbol{y} - E
\boldsymbol{g} - E\tau \dot{\boldsymbol{g}})^T \boldsymbol{B}^{-1} (\boldsymbol{y} - E \boldsymbol{g}- E\tau \dot{\boldsymbol{g}}) + O(\tau^2) \ .
\end{equation}
This leads to the following solution:
\begin{equation}
E = \boldsymbol{w}_{\text{amp}}^T(t_{\mathrm{rel}}) \boldsymbol{y} + O(\tau^2) \quad , \quad 
        \boldsymbol{w}_{\text{amp}}(t_{\mathrm{rel}}) 
        = \frac{ (\dot{\boldsymbol{g}}^T\boldsymbol{B}^{-1}\dot{\boldsymbol{g}}) \boldsymbol{B}^{-1} \boldsymbol{g} -(\boldsymbol{g}^T\boldsymbol{B}^{-1}\dot{\boldsymbol{g}})  \boldsymbol{B}^{-1} \dot{\boldsymbol{g}}}  
        {(\boldsymbol{g}^T \boldsymbol{B}^{-1} \boldsymbol{g})(\dot{\boldsymbol{g}}^T\boldsymbol{B}^{-1}\dot{\boldsymbol{g}}) -(\dot{\boldsymbol{g}}^T\boldsymbol{B}^{-1}\boldsymbol{g})^2 } \ ,
\end{equation}
\begin{equation}
E\tau = \boldsymbol{w}_{\text{time}}^T(t_{\mathrm{rel}}) \boldsymbol{y} + O(\tau^2) \quad ,
        \quad \boldsymbol{w}_{\text{time}}(t_{\mathrm{rel}})
        = \frac{ ({\boldsymbol{g}}^T\boldsymbol{B}^{-1}{\boldsymbol{g}}) \boldsymbol{B}^{-1} \dot{\boldsymbol{g}} -(\boldsymbol{g}^T\boldsymbol{B}^{-1}\dot{\boldsymbol{g}})  \boldsymbol{B}^{-1} {\boldsymbol{g}}}  
        {(\boldsymbol{g}^T \boldsymbol{B}^{-1} \boldsymbol{g})(\dot{\boldsymbol{g}}^T\boldsymbol{B}^{-1}\dot{\boldsymbol{g}}) -(\dot{\boldsymbol{g}}^T\boldsymbol{B}^{-1}\boldsymbol{g})^2 } \ ,
\end{equation}
where $t_{\mathrm{rel}}$ is the time difference between the trigger decision and the first read-out sample, 
see section \ref{sec:reco}. Thus $E$ and $E\tau$ are given by a weighted sum of the discrete measurements 
$y_i$ with the weights for the amplitude, $w_{\text{amp}}(t_{\mathrm{rel}})$, and time shift, 
$w_{\text{time}}(t_{\mathrm{rel}})$, plus $O(\tau^2)$. To reduce $O(\tau^2)$, the fit can be iterated using $g(t_1=t-\tau)$ and the weights $w_{\mathrm{amp/time}}(t_{\mathrm{rel}}+\tau)$~\cite{OF94}. Figure~\ref{fig:Extraction:weights} shows examples of digital filter weights.
% The result for $E$ and $E \tau$ is independent of the number $n$ of FADC measurement used. 
%The first weight $w_{\mathrm{amp/time}}(t_0)$ is plotted as a function of $t_{\mathrm{rel}}$ in the range $[-0.5,0.5[ \ T_{\text{ADC}}$, the second weight in the range $[0.5,1.5[ \ T_{\text{ADC}}$ and so on. 

The expected contributions of the noise to the error of the estimated amplitude and timing only depend on the the shape $g(t)$, and the noise auto-correlation $\boldsymbol{B}$. The corresponding analytic expressions can be found in~\cite{OF94}.

\section{Monte Carlo Studies \label{sec:mc}}

%% Some characteristics of the extractor can only be investigated w
%With the use of Monte-Carlo simulations of signal pulses and noise (for the MAGIC MC simulations, see reference~\cite{Majumdar2005}) %%While in real conditions the charge of the pulse for a given number of incoming Cherenkov photons is Poisson distributed due to the PMT photo-electron statistics, s
%simulated pulses of a specific number of photo-electrons can be generated. 
Pulses of a specific number of photo-electrons can be simulated by using the Monte-Carlo technique to simulate signal pulses and noise (for the MAGIC MC simulations, see reference~\cite{Majumdar2005})
Moreover, using MC, the same pulse can be studied with and without added noise.
%, where the noise level can be deliberately adjusted. 
In the subsequent studies, the Monte Carlo simulation was used to determine, for each of the tested extractors, the following quantities: The bias and the charge 
%and time 
resolution as functions of the input signal charge.% as well as the effect of adding or removing noise to the above quantities.

%\begin{itemize}
%\item The bias as a function of the input signal charge.
%\item The charge resolution as a function of the input signal charge.
%\item The time resolution as a function of the input signal charge.
%\item The effect of adding or removing noise to the above quantities.
%\end{itemize}

For the subsequent studies, the following settings have been used: 

%The LONS has been simulated approximately as in extra-galactic source observation conditions. 
%The electronics noise has been simulated Gaussian at the level measured in data without any correlations between the FADC samples. 
%The intrinsic arrival time spread of the photons was set to be 1\,ns, as expected for $\gamma$-ray showers.

\begin{itemize}
\item{The LONS level in the MC simulations has been set to the value determined from extra-galactic source observation conditions (0.13 photo-electrons per ns, see \cite{GSamlesFADC}).} % about 20\% lower than
\item{The electronics noise has been simulated without any correlations between the FADC samples as a Gaussian distribution with a sigma of 1.6 FADC counts (corresponding to about 0.2 photo-electrons) per FADC slice, roughly at the level measured in data. Note, that in the data the electronic noise introduces a correlation between the FADC samples.}
\item{The intrinsic arrival time spread of the photons was set to be 1\,ns (FWHM of a Gaussian), as expected for $\gamma$-ray showers.}
\item{The conversion factor from photo-electrons to integrated charge over the whole pulse was set to 7.8 FADC counts per photo-electron.}
\item{The relative timing between the trigger and the signal pulse was uniformly distributed over 1 FADC slice.}
\item The total dynamic range of the entire signal transmission chain was set to infinite, thus the detector has been simulated to be completely linear.
%\item Only one inner pixel has been simulated.
\end{itemize}

\subsection{Bias \label{sec:mc:bais}}

\begin{figure}[htp]
\centering
\includegraphics[width=0.485\linewidth]{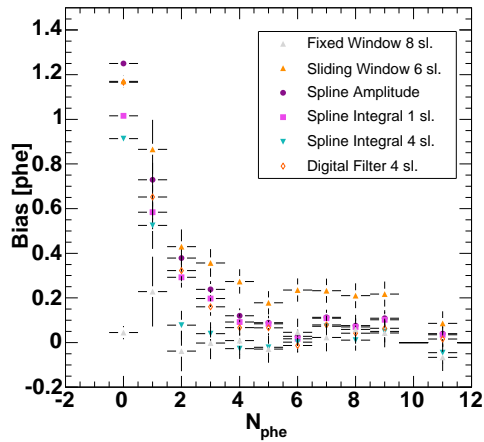}
\caption[Measured Biases from MC simulation]{%
\label{fig:extraction:biases} 
Charge reconstruction bias as a function of the number of generated photo-electrons from MC simulations including  electronic noise plus LONS. Above 12 photo-electrons, the bias vanishes for all signal extractors.}
\end{figure}

%The extracted signal was then converted back to units of photo-electrons, using a fixed conversion factor, and
The signals were simulated with noise and extracted using the different extractor algorithms. For all sliding window algorithms the extraction window was allowed to move 5 FADC slices, independently of its size. For each signal extraction algorithm the average conversion factor between the reconstructed charge in FADC counts and the input number of photo-electrons was determined separately. The signal reconstruction bias was calculated as a function of the simulated number of photo-electrons $N_{\mathrm{sim}}$:
\begin{equation}
B = <\widehat{N}_{\mathrm{rec}} - N_{\mathrm{sim}}>
\end{equation}
Figure~\ref{fig:extraction:biases} shows the results for some 
tested extractors, with different initializations.
As expected, the fixed window extractor does not show any bias up to statistical precision. All other extractors however, do show a bias. Usually, the bias vanishes for signals above 5~photo-electrons, except for the sliding window.
In this latter case, the bias only vanishes for signals above 12~photo-electrons.
% The digital filter and the spline extractors have no bias above 3 photo-electrons and less than 1 photo-electron bias in the case of no signal ($N_{\mathrm{sim}} = 0$).

%%%%%%%%%%%%%%%%%%%%%%%%%%%%%%%%%%%%%%%%%%%%%%%%%%%%%%%%%%%%%%%%%%%%%%%%%%%%%%%%%%%%%%%%%%%%%%%%%%%%

\subsection{Root Mean Square Error \label{sec:mc:rmse}}

\begin{figure}[htp]
\centering
\includegraphics[width=0.485\linewidth]{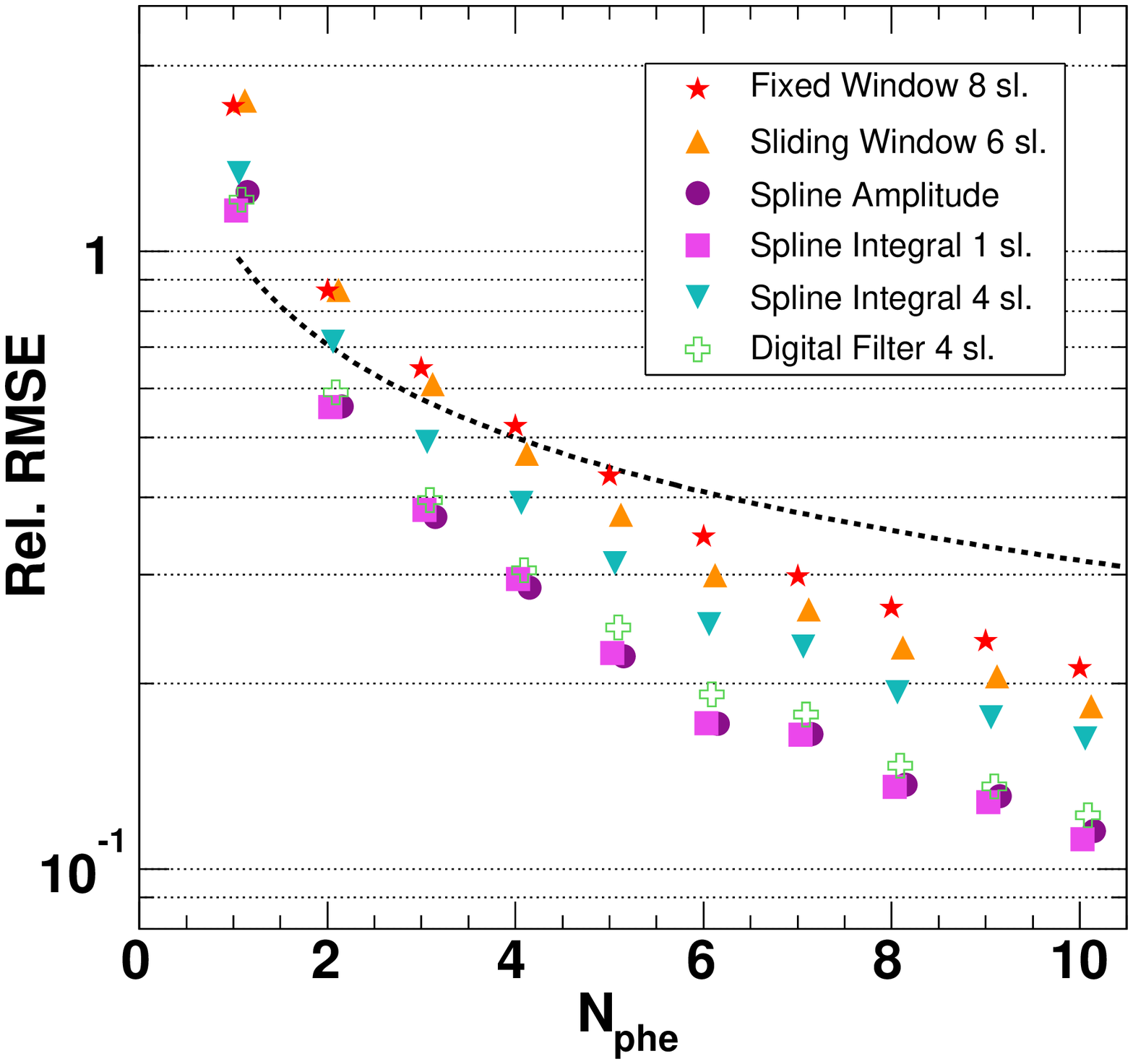}
\includegraphics[width=0.485\linewidth]{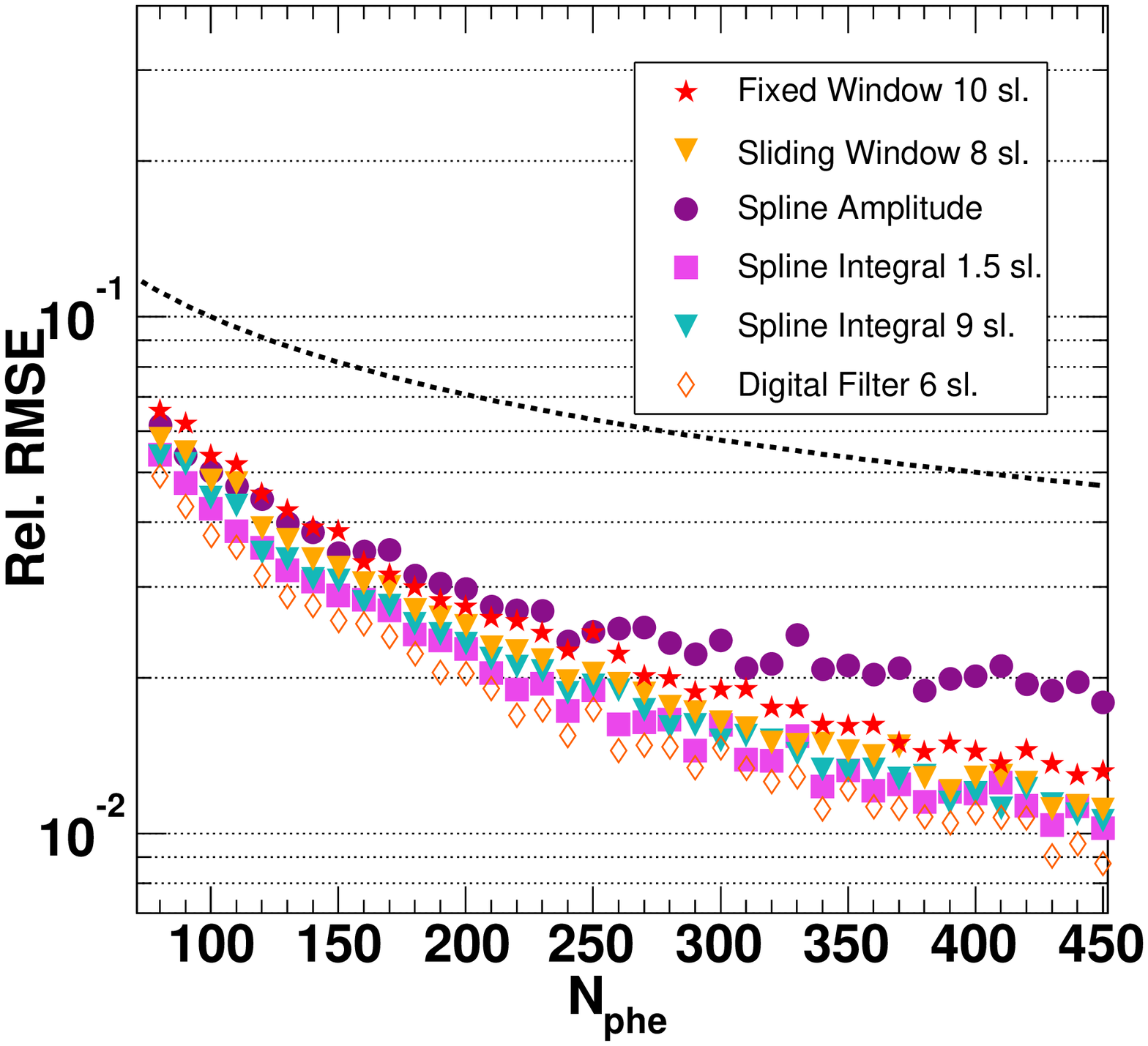}
\caption[Relative Root-Mean-Square-Error from MC simulation]{%
\label{fig:extraction:resolutions} 
Relative \textit{RMSE} as a function of the number of generated photo-electrons from MC simulations including fully simulated electronic noise plus LONS. Left: high gain, right: low gain. The black dashed line shows the square root of the relative variance of the incoming numbers of photo-electrons, note, that the PMT introduces an additional excess noise \cite{mirzoyanlorenz}. The best results are obtained with the digital filter or a spline integrating 1 FADC slice.}
\end{figure}

In order to obtain the precision of a given extractor, we calculated the relative \textit{RMSE}:
\begin{equation}
\mathrm{Rel.}~\textit{RMSE} = \frac{1}{N_{\mathrm{sim}}} \sqrt{ \mathrm{Var}[\widehat{N}_{\mathrm{rec}}] + B^2} \ .
\end{equation}
Figure~\ref{fig:extraction:resolutions} shows the relative \textit{RMSE} for the high gain and low gain parts separately. Also the square root of the relative variance of the number of simulated photo-electrons ($\sqrt{1/N_{\mathrm{sim}}}$) is shown, which corresponds to the intrinsic fluctuations of the signal from air showers, following Poissonian statistics. Note, that the PMT introduces an additional excess noise \cite{mirzoyanlorenz}, which is on average 18\% of the Poissonian fluctuations for the MAGIC PMTs. For all extractors the variance of the reconstructed signal is dominated by noise and only slowly increases with rising signals due to mis-reconstruction of the signal pulse itself. Therefore, the relative \textit{RMSE} is proportional to $1/N_{\mathrm{phe}}$. For small numbers of photo-electrons, extractors with small extraction windows or the digital filter yield the smallest values of \textit{RMSE}, but the difference is only important below about 5 photo-electrons. Above that value, the curves for all extractors have crossed the black line, i.e. they are more precise than the intrinsic fluctuations of the signal. This is also true for the entire low gain extraction range. 
%Therefore, $\mathrm{Rel.}~\textit{RMSE} \propto 1/N_{\mathrm{phe}}$, whereas the relative size of the Poisson fluctuations of the signal is $1/{\sqrt{N_{\mathrm{phe}}}}$.
The best results are obtained with the digital filter or a spline integrating 1 FADC slice. 
%For the relatively narrow high gain pulse the spline amplitude yields comparable results, nevertheless for the wider low gain pulse the spline amplitude is subject to larger fluctuations and gives worse results compared to the digital filter and the spline integral.

\section{Pedestal Extraction \label{sec:pedestals}}

The pedestal is the average FADC count for the signal baseline (no input signal). 
To determine the pedestal setting off-line, dedicated pedestal runs are used, during which the 
MAGIC read-out is triggered randomly. The fluctuations of the signal baseline are due 
to electronics noise and LONS fluctuations. Thus the pedestal RMS is a measure for the 
total noise level. 
%It can be completely described by the noise-autocorrelation matrix $\boldsymbol{B}$, see equation (\ref{eq:Extraction:autocorr}). $\boldsymbol{B}$ is independent of the signal extraction algorithm.

%By applying the signal extractor with a global extraction window to pedestal events, allowing it to ``slide'' and maximize the encountered signal, the bias $B$ and the root-mean-square-error \textit{RMSE} for the case of no signal ($N_{phe}=0$) can be determined. 

\begin{landscape}
%\rotatebox{90}{%
\begin{table}[htp]
\centering
{\scriptsize
\begin{tabular}{|c|c|c|c|c|c|c|c|c|c|c|c|c|c|c|}
%\hline
%\hline
%\multicolumn{15}{|c|}{\large{Statistical Parameters for $N_\mathrm{phe}=0$} \rule{0mm}{6mm} \rule[-2mm]{0mm}{6mm} } \\
\hline
\hline
& \multicolumn{3}{|c|}{Closed camera} & \multicolumn{3}{|c|}{MC simulation}  & \multicolumn{4}{|c|}{Extra-galactic LONS}  & \multicolumn{4}{|c|}{Galactic LONS} \ruu\ruo \\
\hline
\hline
 Name  & $\sqrt{\mathrm{Var}[N_{rec}]}$ & $B$ & \textit{RMSE} 
       & $\sqrt{\mathrm{Var}[N_{rec}]}$ & $B$ & \textit{RMSE} 
       & $\sqrt{\mathrm{Var}[N_{rec}]}$ & $B$ & \textit{RMSE} & \multicolumn{1}{c|}{$N_{\mathrm{phe}}^{\mathrm{thres.}}$} 
       & $\sqrt{\mathrm{Var}[N_{rec}]}$ & $B$ & \textit{RMSE} & \multicolumn{1}{c|}{$N_{\mathrm{phe}}^{\mathrm{thres.}}$} \ruo \ruu \\
\hline                                                     
\hline                                                     
Fixed Win. 8  & 1.2 & \textcolor{red}{\bf 0.0} & 1.2  
              & 2.1 & \textcolor{red}{\bf 0.0} & 2.1   
              & 2.5 & \textcolor{red}{\bf 0.0} & 2.5 & 7.5 
              & 3.0 & \textcolor{red}{\bf 0.0} & 3.0 & 9.0 \ruo \ruu \\   
\hline                                                     
%% Slid. Win. 1  & 0.4 & 0.4 & 0.6
%%               &     &     &      
%%               & 1.2 & 1.3 & 1.8 & 4.9 
%%               & 1.4 & 1.5 & 2.0 & 5.7 \ruo\\
Slid. Win. 2  & 0.5 & 0.4 & 0.6
              & 1.1 & 1.0 & 1.5    
              & 1.4 & 1.2 & 1.8 & 5.4 
              & 1.6 & 1.5 & 2.2 & 6.1 \ruo \\
Slid. Win. 4  & 0.8 & 0.5 & 0.9 
              & 1.4 & 1.1 & 1.8    
              & 1.9 & 1.2 & 2.2 & 6.9 
              & 2.3 & 1.6 & 2.8 & 7.5 \\
Slid. Win. 6  & 1.0 & 0.4 & 1.1 
              & 1.8 & 1.0 & 2.1    
              & 2.2 & 1.1 & 2.5 & 7.7 
              & 2.7 & 1.4 & 3.0 & 9.5 \\
Slid. Win. 8  & 1.3 & 0.4 & 1.4 
              & 2.1 & 0.8 & 2.2    
              & 2.5 & 1.0 & 2.7 & 8.5 
              & 3.2 & 1.4 & 3.5 & 10.0 \ruu \\
\hline                                                                              
Spline Amp.   & \textcolor{red}{\bf 0.4} & 0.4 & 0.6  
              & 1.1 & 1.1 & 1.6    
              & 1.2 & 1.3 & 1.8 & 4.9 
              & 1.4 & 1.6 & 2.1 & 5.8 \ruo \\
\textcolor{red}{\bf Spline Int. 1} & \textcolor{red}{\bf 0.4} & 0.3 & \textcolor{red}{\bf 0.5} 
              & 1.1 & 0.8 & 1.4    
              & 1.2 & 1.0 & 1.6 & 4.6 
              & \textcolor{red}{\bf 1.3} & 1.3 & 1.8 & 5.2 \\
Spline Int. 2 & 0.5 & 0.3 & 0.6  
              & 1.2 & 0.9 & 1.5    
              & 1.4 & 0.9 & 1.7 & 5.1 
              & 1.6 & 1.2 & 2.0 & 6.0 \\
Spline Int. 4 & 0.7 & \textcolor{red}{\bf 0.2 } & 0.7  
              & 1.3 & 0.8 & 1.5    
              & 1.7 & \textcolor{red}{\bf 0.8}  & 1.9 & 5.3 
              & 2.0 & 1.0 & 2.2 & 7.0 \\
Spline Int. 6 & 1.0 & 0.3 & 1.0 
              & 1.7 & 0.8 & 1.9    
              & 2.0 & \textcolor{red}{\bf 0.8} & 2.2 & 6.8 
              & 2.5 & \textcolor{red}{\bf 0.9} & 2.7 & 8.4 \ruu \\
\hline                                                                              
\textcolor{red}{\bf Dig. Filt. 4} & \textcolor{red}{\bf 0.4} & 0.3 & \textcolor{red}{\bf 0.5} 
              & 1.0 & 1.3 & 1.6    
              & \textcolor{red}{\bf 1.1} & 0.9 & \textcolor{red}{\bf 1.4} & \textcolor{red}{\bf 4.2} 
              & \textcolor{red}{\bf 1.3} & 1.1 & \textcolor{red}{\bf 1.7} & \textcolor{red}{\bf 5.0}\ruo \\
Dig. Filt. 6  & 0.5 & 0.4 & 0.6  
              & 1.1 & 1.3 & 1.7    
              & 1.3 & 1.3 & 1.8 & 5.2 
              & 1.5 & 1.5 & 2.1 & 6.0 \ruu \\
\hline
\hline
\end{tabular}
}
\caption[Bias and \textit{RMSE} from pedestal events for various signal extractors]{%
\label{tab:Extraction:pedestalbias}
The statistical parameters square root of reconstructed signal variance, bias,  \textit{RMSE} and $N_{\mathrm{phe}}^{\mathrm{thres.}}$ for the tested signal extractors, applied to pedestal events. All units are in reconstructed numbers of photo-electrons, statistical uncertainty: about 0.1 photo-electrons. The extractors yielding the smallest values for each column are marked in red.
%The first line shows the results of the smallest robust fixed--window extractor for reference. 
}
\end{table}
\end{landscape}

By applying the signal extractor to pedestal events, the bias $B$ and the \textit{RMSE} for the case of no signal ($N_{\mathrm{phe}}=0$) can be determined. 
Table~\ref{tab:Extraction:pedestalbias} shows the bias, the square root of the variance $\mathrm{Var}[N_{\mathrm{rec}}]$ and the root-mean-square error for randomly triggered pedestal events with closed camera and sample observations outside the Galactic plane (``extra-galactic LONS'', $~0.13$ photo-electrons/ns) as well as within the Galactic plane (``galactic LONS''). In addition, table~\ref{tab:Extraction:pedestalbias} shows the corresponding values from MC simulations. The $\mathrm{Var}[N_{\mathrm{rec}}]$ of the MC simulations is slightly lower then in the sample observations outside the Galactic plane, although the simulated LONS level in the MC simulations has been adjusted to the ``extra-galactic LONS''. This is in part due to neglecting the correlation of the FADC slices from the electronic noise, see section \ref{sec:mc}.
%In the MC a slightly lower level of LONS has been simulated then in the sample observations outside the Galactic plane.
%In this example, e 
Every extractor window had the freedom to move 5 FADC slices, i.e. the global window size was fixed to five plus the extractor window size. 
%The first line shows the resolution of the smallest existing robust fixed window algorithm in order to give the reference value of 2.5 and 3 photo-electrons RMS for an extra-galactic and a galactic star-field, respectively. 

% The \textit{RMSE} reaches its lowest values if the smallest sliding window sizes are used.
One can see that the bias typically decreases and the variance increases with increasing sliding window size, except for the digital filter. 
%The fixed window algorithm integrating eight FADC slices shows the worst \textit{RMSE} of 2.5 and 3 photo-electrons for an extra-galactic and a galactic star-field, respectively. Nevertheless, it has no bias. 
The extractor with the smallest \textit{RMSE} is the digital filter fitting 4 FADC slices with an \textit{RMSE} of 1.4 and 1.7  photo-electrons for an extra-galactic and a galactic star-field, respectively. 

% All sliding window extractors have a smaller \textit{RMSE} than the RMS of the signal from the fixed window reference extractor. This shows that the global error of the sliding window extractors is smaller than the one of the fixed window extractor even if the first ones have a bias.

%, e.g. for the image cleaning,  

In the so-called image cleaning \cite{Fegan1997} only the camera pixels above a certain charge threshold are used for the image parameterization \cite{Hillas_parameters}. The charge threshold is adjusted such that the probability of being a noise fluctuation does not exceed a certain value. For the sake of comparison, a typical value of 3$\,\sigma$ (0.3\% probability) was chosen here and that number approximated with the formula: 
\begin{equation}
N_{\mathrm{phe}}^{\mathrm{thres.}} \approx B + 3 \cdot \sqrt{V} \ .
\label{eq:Extraction:nphethres}
\end{equation}
%
%where $V$ is the variance of the extracted signal (eq.~\ref{eq:def:r}). 
$N_{\mathrm{phe}}^{\mathrm{thres.}}$ is shown in the 11$^\mathrm{th}$ and 15$^\mathrm{th}$ column of table~\ref{tab:Extraction:pedestalbias}. Most of the sliding window algorithms yield a smaller signal threshold than the fixed window, although the former ones have a bias. 

The lowest threshold of only 4.2~photo-electrons for the extra-galactic star-field and 5.0~photo-electrons for the galactic star-field is obtained with the digital filter fitting 4 FADC slices. This is almost a factor 2 lower than the fixed window results. 
%Also the spline integrating 1 FADC slice yields comparable results. 
%In addition, table~\ref{tab:Extraction:pedestalbias} shows the corresponding values from MC simulations. In the MC a slightly lower level of LONS has been simulated.
%Here, one has to take into account the slightly lower level of simulated LONS. Therefore, about 20\% lower values of \textit{RMSE} are expected than in the case of real data.

\section{Calibration \label{sec:calibration}}

In this section, tests are described which were performed using light pulses of different 
color, shape and intensity produced by the MAGIC LED calibration pulser 
system~\cite{MAGIC_calibration}. Such a system is able to provide fast light pulses 
of 2--4\,ns FWHM, with intensities ranging from 3 to more than 600 photo-electrons in one 
inner PMT of the MAGIC camera. These pulses can be produced in three colors: 
{\textit {\bf green, blue}} and {\textit{\bf UV}}. Table~\ref{tab:pulsercolors} 
lists the available colors and intensities.

\begin{table}[htp]
\centering
\begin{tabular}{|c|c|c|c|c|c|c|}
%\hline
%\hline
%\multicolumn{7}{|c|}{The possible pulsed light colors} \\
\hline
\hline
Color &  Wavelength & Spectral Width & Min. No. &  Max. No. & Secondary & FWHM \\
      & [nm]         & [nm]           &  Phe's   &  Phe's    & Pulses  &  Pulse [ns]\\
\hline
Green &  520      & 40      & 6          &  120      & yes  & 3--4  \\
\hline
Blue &  460       & 30      & 6          &  600      & yes  & 3--4 \\
\hline
UV   &  375       & 12      & 3          &  50       & no   & 2--3 \\ 
\hline
\hline
\end{tabular}
\caption{The pulser colors available from the calibration system}
\label{tab:pulsercolors}
\end{table}

% and 
%Figure~\ref{fig:pulseexample} shows examples of the smallest and largest intensity calibration pulses as recorded by the FADCs. 
Whereas the pulse shape of the UV LEDs is very stable from event to event, the green and blue LED pulses can show smaller secondary pulses about 10--40\,ns after the main pulse. Note, that the UV-pulses are only available in intensities which do not saturate the high gain read-out channel. However, the brightest combination of (blue) light pulses easily saturates all high gain channels of the camera, but does not saturate the low gain read-out.
%\par
%The tests comprise three items:

%\begin{enumerate}
%\item Number of photo-electrons: The reconstructed number of photo-electrons should be independent of the signal extraction algorithm used.
%\item Robustness tests: These tests investigate, for the different extractors, the stability of the reconstructed charges in FADC counts and number of photo-electrons (after calibration) in case of variations of the pulse form for the different extractors. 
%\item Time resolution: These tests show the time resolution for different intensities and colors.
%\end{enumerate}

%\begin{figure}[htp]
%\centering
%\includegraphics[width=0.37\linewidth]{1LedUV_Pulse_Inner.eps}
%\includegraphics[width=0.37\linewidth]{1LedUV_Pulse_Outer.eps}
%\includegraphics[width=0.37\linewidth]{23LedsBlue_Pulse_Inner.eps}
%\includegraphics[width=0.37\linewidth]{23LedsBlue_Pulse_Outer.eps}
%\caption[Example Calibration Pulses Lowest and Highest Intensity]{%
%\label{fig:pulseexample}
%Example of a calibration pulse from the lowest (top) and highest (bottom) available 
%mono-chromatic intensity. Left: a typical inner pixel, right: a typical outer pixel. Note that in the upper plots, the pulse height fluctuates much more than suggested from these pictures. Especially, a zero-pulse is also possible. In the lower plots, the (saturated) high gain channel is visible at early times, while from FADC slice 20 on, the delayed low gain pulse appears. }
%\end{figure}

\subsection{Number of Photo-electrons \label{sec:Extraction:nphe}}

The mean number of photo-electrons $<\widehat{N}_\mathrm{phe}>$ was calculated for a sequence of calibration pulses of same intensity, following the excess noise factor method~\cite{mirzoyanlorenz} and using different signal extractor algorithms. If the signals are extracted correctly,  $<\widehat{N}_\mathrm{phe}>$ should be independent of the signal extractor.

In our case, an additional complication arises from secondary pulses of the green and blue colored light pulses, which may introduce a dependence of  $<\widehat{N}_\mathrm{phe}>$ on the extraction time-window size (recall figure~\ref{fig:pulse_shapes}). For the standard MAGIC calibration procedure \cite{MAGIC_calibration} only UV calibration pulses are used.
% The signal extractors will have to be grouped into those which are affected by the secondary pulses and those being immune to this effect. 

\begin{figure}[htp]
\centering
\includegraphics[width=0.485\linewidth]{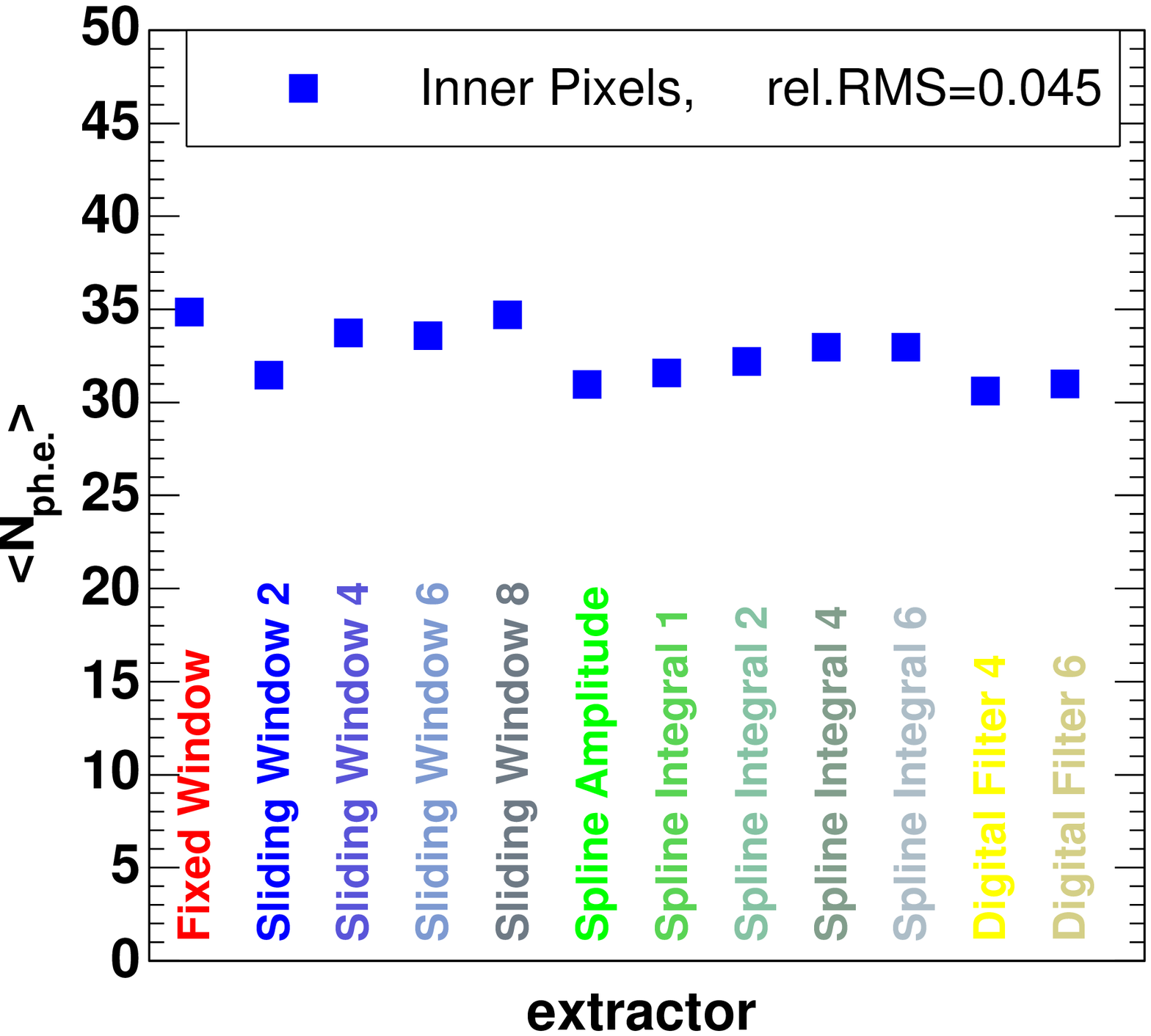}
\includegraphics[width=0.485\linewidth]{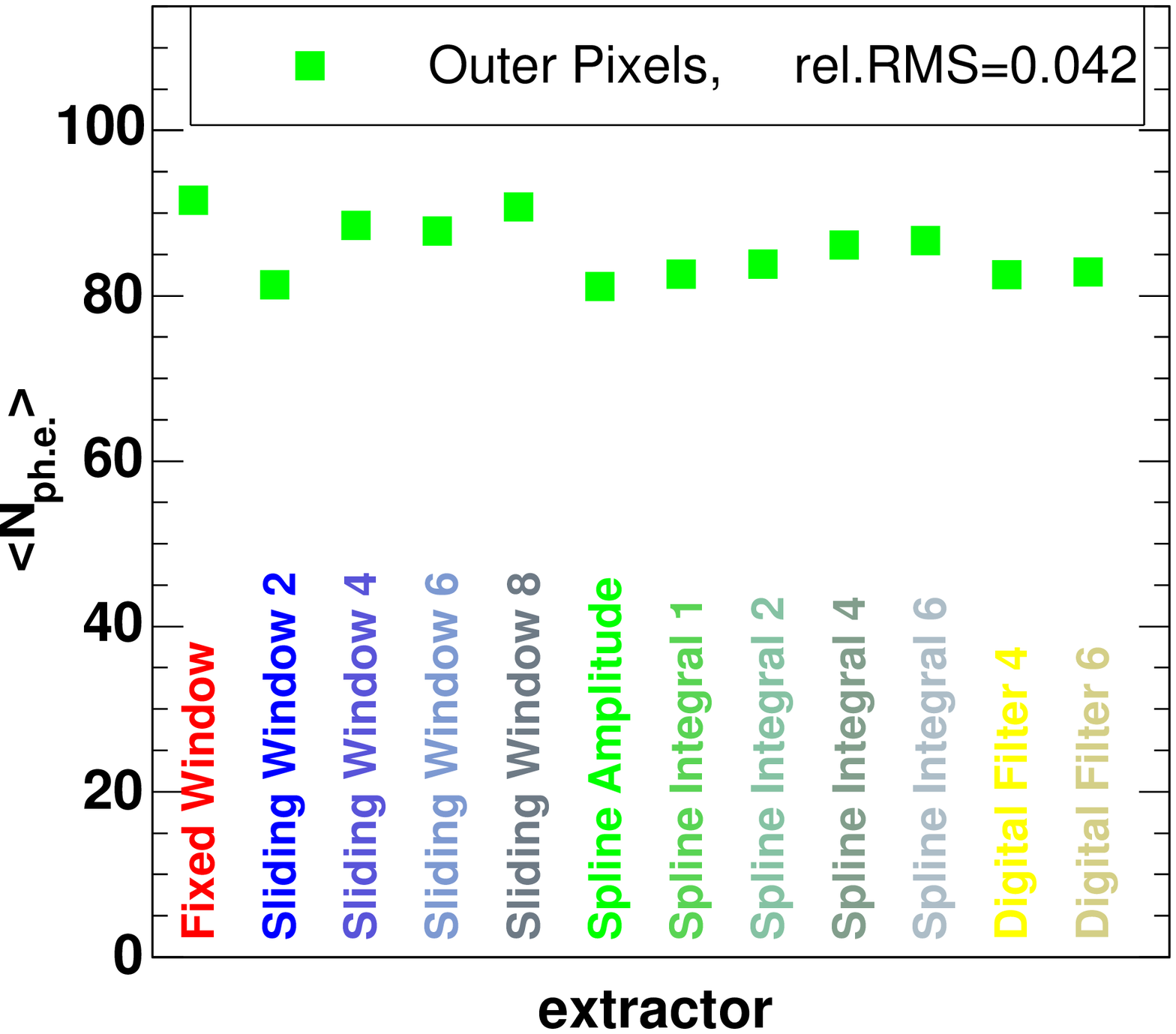}
\caption[Reconstructed Number of Photo-electrons standard UV calibration pulse]{%
$<\widehat{N}_\mathrm{phe}>$ from the standard calibration pulse, emitted by 10 UV LEDs, reconstructed with different signal extractors. Left: inner pixels, right: outer pixels. The statistical errors are smaller than the marker size.
\label{fig:phe:10ledsuv} }
\end{figure}

Figure~\ref{fig:phe:10ledsuv} shows $<\widehat{N}_\mathrm{phe}>$ for the standard UV calibration pulse. The results differ by less than 5\%, which results in an additional systematic error to the absolute energy scale of the reconstructed events. Note, that the total systematic error of the absolute energy scale was estimated to be 16\% \cite{MAGIC_Crab}. 
%VHE $\gamma$-ray flux level, which is in total estimated to be 35\% !!! look at Crab paper !!! \cite{MAGIC_Crab}. 
A small increase of $<\widehat{N}_\mathrm{phe}>$ for an increasing window size can be observed. This may be due to the intrinsic time structure of the calibration pulse.

% tails in the intrinsic time structure of the calibration pulse.
%% \par
%% On the contrary, a typical green pulse yields time-window size dependent numbers of photo-electrons, as shown 
%% in figure~\ref{fig:phe:5ledsgreen}. The obtained spread in number of photo-electrons lies around 15\%, due to that 
%% dependency. Applying corrections to the variance does not reduce the spread considerably. 
%% The digital filter reconstructs thereby the same 
%% number of photo-electrons as the other extractors using small window sizes, i.e. the secondary pulses are efficiently 
%% filtered out.

%% A similar behavior can be seen in the third example of an intense blue calibration pulse 
%% (figure~\ref{fig:phe:20ledsblue}): A spread of about 12\% is obtained, independently whether corrections are 
%% applied or not. Here, the digital filter yields results comparable to medium window sizes of other extractors, 
%% most probably because the secondary pulses influence already the overall pulse form, due to the limited bandwidth of 
%% the passive delay line.

The peak-to-peak variation of the conversion factor between FADC counts and number of 
photo-electrons for the different intensities is below 10\% \cite{markus} for all extractors. The corresponding non-linearity is due to the intrinsic non-linearity of the MAGIC signal chain and a possible non-linear signal extraction.

\subsection{Robustness Tests \label{sec:Extraction:robustness}}

%Typically, two sources of degradations of the signal extraction quality are of concern for the MAGIC data: {\bf variations of the pulse shape} and possible {\bf early or late pulse positions within the recorded FADC samples}. 

Possible {\bf variations of the pulse shape} may degrade the signal extraction quality of the MAGIC data.
Variations of the pulse form have a physical reason: average Cherenkov pulses from 
hadronic showers are usually broader than those from electromagnetic cascades. 
Additionally there are differences between the pulse form of calibration pulses and 
those of cosmic pulses. These variations affect mainly those signal extractors which 
integrate only parts of a pulse or perform fits to a sample pulse form. In order to 
quantify the magnitude of the effect, table~\ref{tab:Extraction:pulseformdep} lists the 
fraction of the pulse which is contained in typical time windows around the pulse 
maximum, for various pulse forms. While the amplitude extraction or integration of 
only 1~FADC slice around the maximum yield differences as high as 10\% (cosmic -- UV), 
the error is reduced to about 3\% if four FADC slices are being integrated. Further deviations, characterized by the blue LED calibration pulse or the MC pulse, yield an even stronger discrepancy.

\begin{table}[htp]
\centering
\begin{tabular}{|c||cccc|cc|}
%\hline
%\hline
%\multicolumn{7}{|c|}{\large Pulse Form Dependency of Integration Windows} \ruu\ruo \\
\hline
\hline
Window Size     & \multicolumn{4}{c|}{High Gain} & \multicolumn{2}{c|}{Low Gain} \ruo \\
(FADC slices    & \small{MC} & \small{Cosmic} & \small{Calib. UV} & \small{Calib. Blue} 
                &              \small{Cosmic} &                     \small{Calib. Blue}\\
around maximum) & \multicolumn{6}{c|}{(percentage of complete pulse integral)} \ruu \\
\hline
Amplitude         & 5.4 & 5.0   & 4.5 & 4.1  & 3.4 & 2.5 \ruo\\
1 slice         & 54  & 50    & 46  & 41   & 35  & 27 \\
2 slices        & 78  & 76    & 71  & 66   & 60  & 48 \\
4 slices        & 97  & 98    & 95  & 89   & 90  & 82 \ruu \\
\hline
\end{tabular}
\caption[Pulse Form Dependency of Integration Windows]{%
\label{tab:Extraction:pulseformdep}
Pulse form dependency of integration windows: Shown is the fraction of the signal 
(in percent of the complete pulse integral), contained in different time windows around 
the pulse maximum for different pulse shapes. In the case of the first line ``Amplitude'', the signal amplitude has simply been divided by the complete pulse integral (arbitrary units).}
\end{table}

\begin{figure}[htp]
\centering
\includegraphics[width=0.485\linewidth]{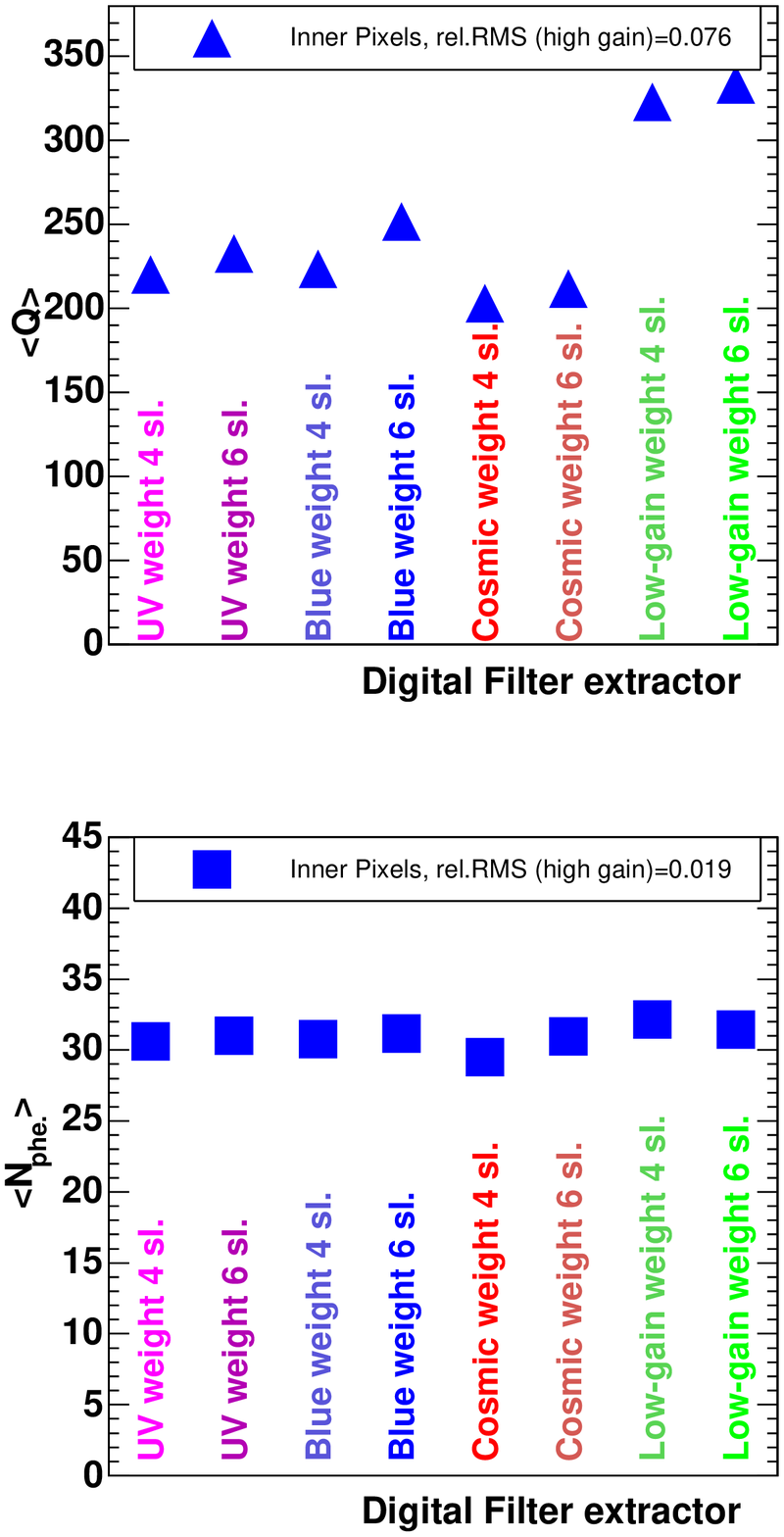}
\includegraphics[width=0.485\linewidth]{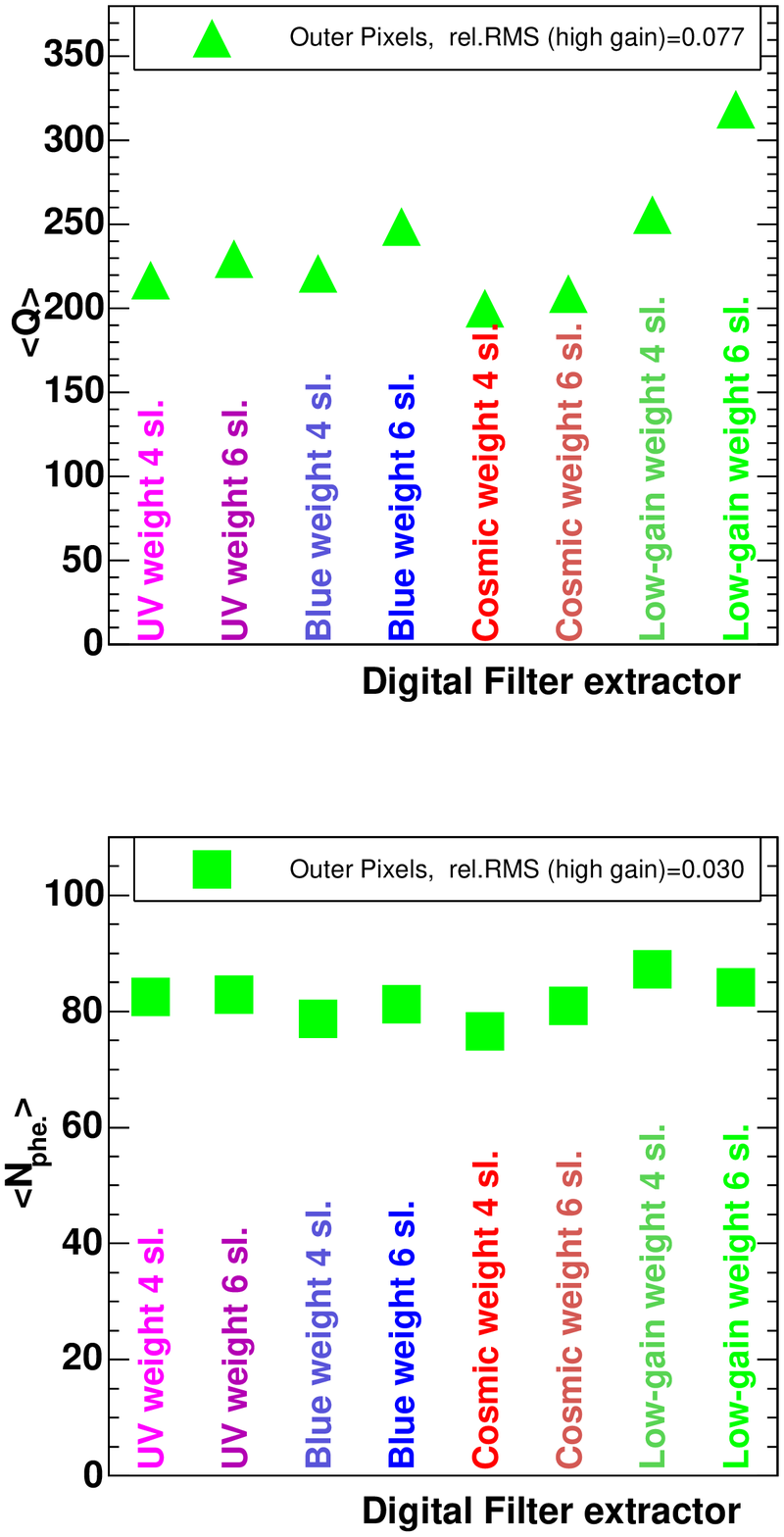}
\caption[Reconstructed Number of Photo-electrons different pulse forms]{%
Mean reconstructed charge in FADC counts (top) and $<\widehat{N}_\mathrm{phe}>$ (bottom) from a standard calibration pulse reconstructed with a digital filter using different weight functions (computed for UV and blue calibration pulses, Cherenkov pulses and the low gain). Left: inner pixels, right: outer pixels. The relative RMS was calculated for the first six (high gain) pulse forms. The statistical errors are smaller than the marker size.
\label{fig:Extraction:10ledsuv:dftest} }
\end{figure}

The digital filtering method assumes a constant signal shape to compute the weight 
functions. In fact, all pixels are assumed to have the same average signal shape and 
the same weights are used for all pixels. In order to test the robustness of the digital 
filtering method with respect to deviations of the actual pulse shape from the assumed 
pulse shape, the standard UV calibration pulse was extracted using different 
weight functions (computed for UV and blue calibration pulses, Cherenkov pulses and the 
low gain).
%, i.e. fitted to different pulse forms. 
The results are displayed in figure~\ref{fig:Extraction:10ledsuv:dftest} showing 
variations of about 8\% in the reconstructed signal for typical pulse form variations (blue, UV and cosmic weights)
and 3\% in  $<\widehat{N}_\mathrm{phe}>$ after calibration using the same weights. Note, photons from a $\gamma$-ray shower (from a weighted spectrum) arrive within (2-2.5)~ns \cite{muon_rejection}.

In conclusion, an event-to-event variation of the pulse shape may cause a charge 
reconstruction error of up to 10\% for all signal extractors which do not integrate the 
entire pulse. The size of this error decreases with increasing integration window size. 
For the digital filter this event-to-event variation of the pulse shape may lead to an error of the reconstructed signal of up to 8\%. A systematic difference in the pulse shape causes an error of up to 3\% (after calibration) in the case of the digital filter.

%A second kind of reconstruction errors may be due to the spread of pulse positions within the recorded FADC samples. There is a physical spread of about 2.5 FADC samples RMS, due to the time spread of showers and to the trigger jitter of about one FADC slice. The read-out has to be adjusted such that the whole pulse is still inside the digitization window.

%If the read-out is adjusted such that the {\bf\textit{mean pulse position}} is too early or too late, a considerable part of the signals cannot be extracted any more because the pulses may reach out of the registered FADC window.

% An additional offset of about 5\:ns occurred in one trigger cell, due to hardware problems, which were cured only in September, 2005 and affect all data taken before. Taking together these numbers, the position of a pulse from an air shower may vary within 7--8 FADC slices, within one data run. If the read-out is adjusted such that the {\bf\textit{mean pulse position}} is too early or too late, a considerable part of the signals cannot be extracted any more because the pulses may reach out of the registered FADC window. This problem affects above all extractors which use large extraction windows, especially the standard digital filter fitting the low gain pulse to a sample of 6 FADC slices. 

%%%%%%%%%%%%%%%%%%%%%%%%%%%%%%%%%%%%%%%%%%%%%%%%%%%%%%%%%%%%%%%%%%%%%%%%%%%%%%%%%%%%%%%%%%%%%%%%

\subsection{Time Resolutions \label{sec:cal:timeres}}

%Since the calibration system does not deliver a precise enough absolute arrival time stamp, a measurement of the relative arrival time $\delta t$ has to be made
The calibration light pulses can be used to test the time resolution of signal extractors. Thereby, the arrival time difference $\delta t$ is measured for every channel, with respect to a reference channel: 
\begin{equation}
\delta t_i = t_i - t_\mathrm{ref}~,
\end{equation}
where $t_i$ denotes the reconstructed arrival time of pixel number $i$ and $t_\mathrm{ref}$ the reconstructed arrival time of a reference pixel. Using a calibration run of a fixed number of calibration pulses, the mean and RMS of the distribution of $\delta t_i$ for a given pixel can be computed. The RMS is a measure of the combined time resolutions of pixel $i$ and the reference pixel. Assuming that the photomultipliers and read-out channels are of the same kind, an approximate time spread of pixel $i$ is obtained from the sigma of a Gaussian fit to the distribution of the time differences $\delta t_i$:
\begin{equation}
\Delta t_i \approx \sigma(\delta t_i)/\sqrt{2}~. %\mathrm{RMS}
\end{equation}
\begin{figure}[htp]
\centering
\includegraphics[width=0.32\linewidth]{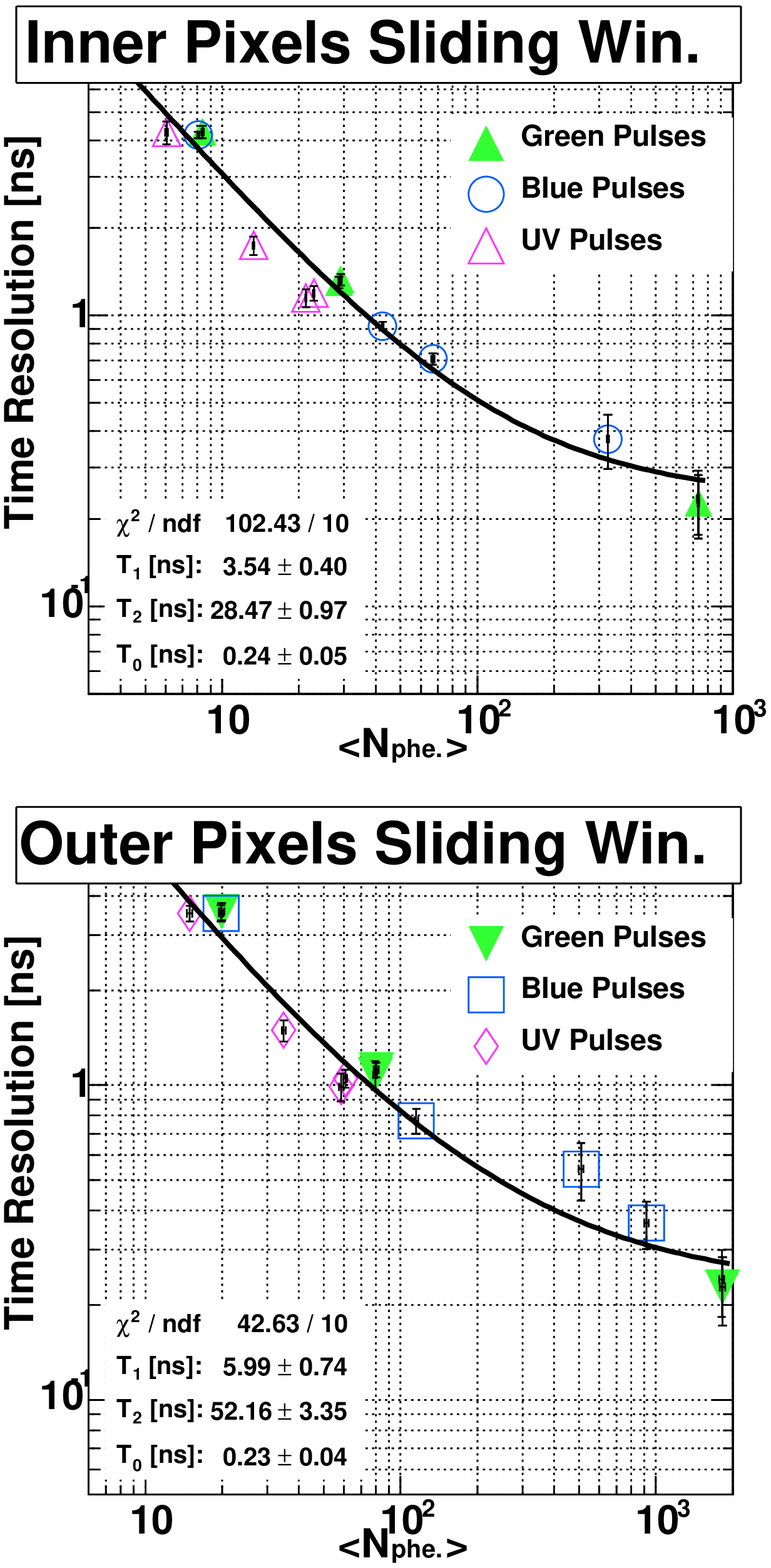}
\includegraphics[width=0.32\linewidth]{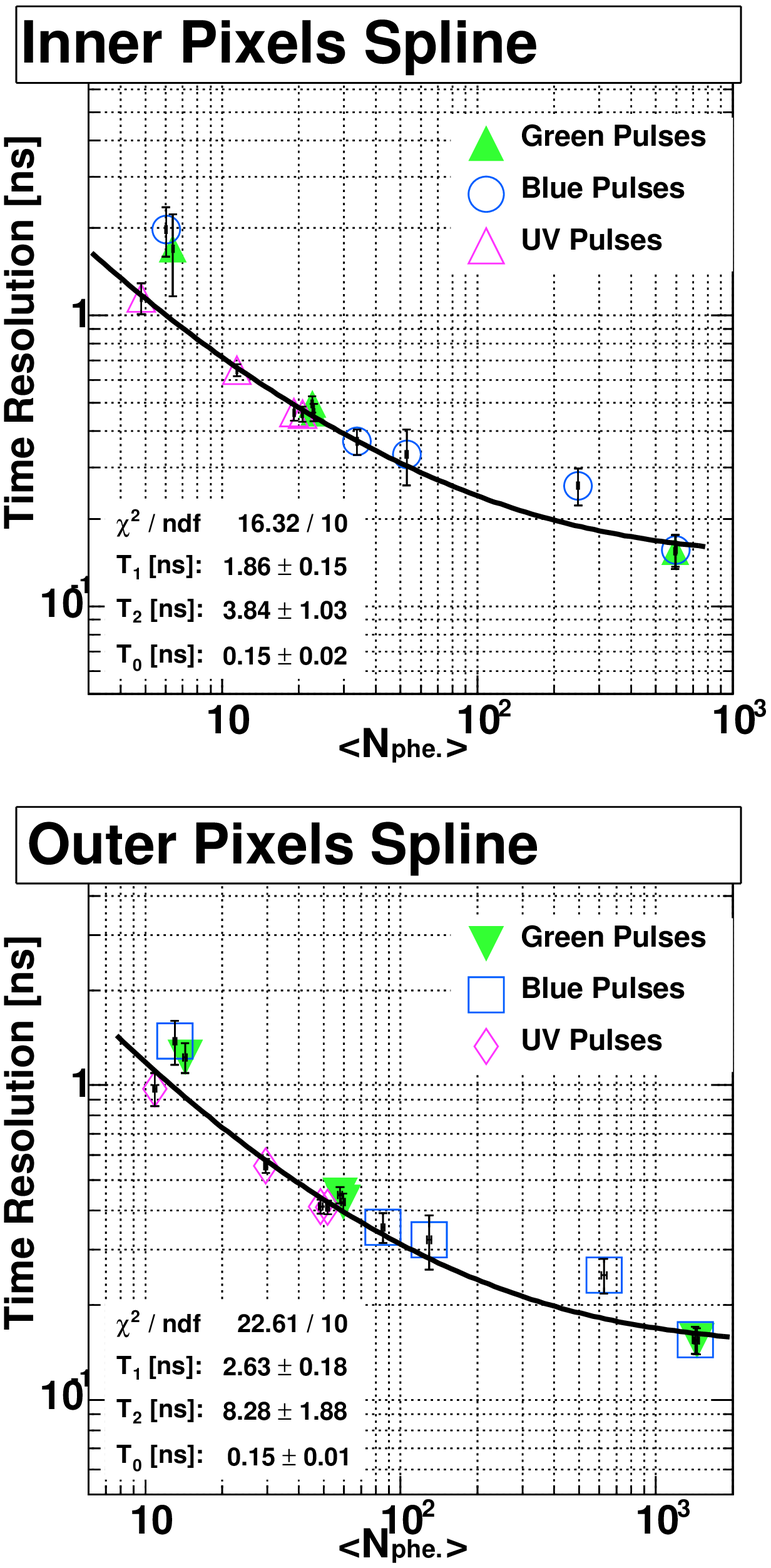}
\includegraphics[width=0.32\linewidth]{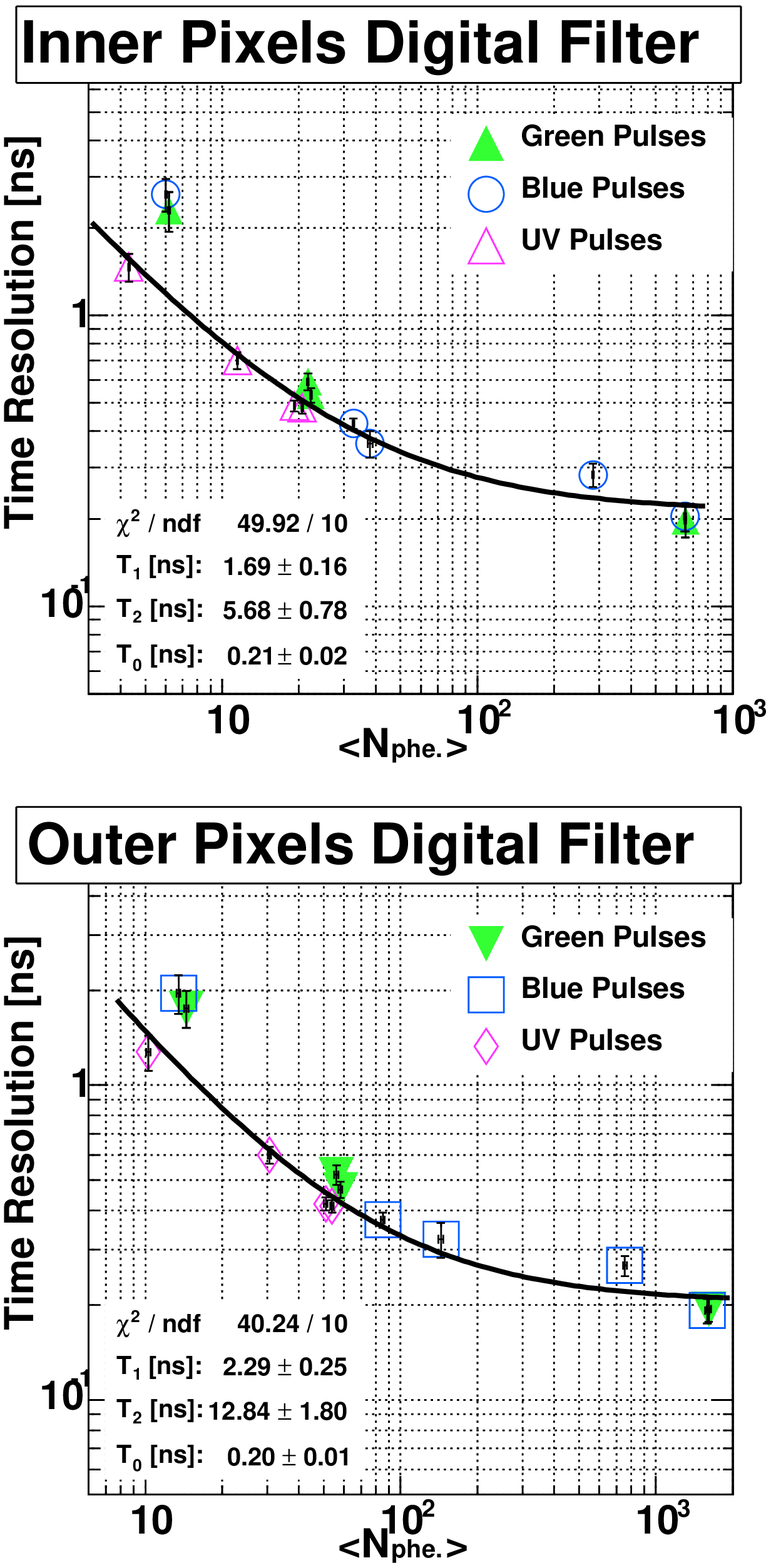}
\caption[Arrival Time Resolutions]{%
\label{fig:Extraction:timedep}
Reconstructed mean arrival time resolution as a function of the extracted mean number of 
photo-electrons for three different extractors: the amplitude-weighted sliding window 
with a window size of 6~slices (left), the half-maximum searching spline (center), and
the digital filter with correct pulse weights over 6 slices (right). 
Error bars denote the spread (RMS) of time resolutions of the investigated channels. 
The marker colors show the applied pulser color, except for the last (green) point 
where all three colors were used. The lines are a fit using equation 
(\ref{eq:Extraction:timefit}), see text for details. The best fit parameters are shown 
as an inset.}
\end{figure}
%\end{landscape}

Figure~\ref{fig:Extraction:timedep} shows the obtained average time resolutions 
$<\Delta t_i>$ as a function of $<\widehat{N}_\mathrm{phe}>$ for various calibration runs taken with different colors and light intensities for the telescope pointing outside the Galactic plane (``extra-galactic LONS'', ~0.13 photo-electrons/ns). 
Three time extractors were used: a Sliding Window of 6 FADC slices with amplitude-weighted time, the Cubic Spline with the position of the half-maximum at the rising edge of the pulse as arrival time and the digital filter. 
%Note, that the time spread decreases with increasing signal charge, as expected, and a
Note, that a time resolution of better than 1\,ns can be obtained for all pulses above a 
threshold of 5~photo-electrons. For the largest signals, a time resolution as good as 
200\,ps can be obtained. In order to understand the exact behavior of the time resolution,
we briefly review the main contributions:

\begin{enumerate}
\item The intrinsic arrival time spread of the photons on the PMT: This time spread can be estimated roughly by the intrinsic width $\delta t_{\mathrm{IN}}$ of the input light pulse. The resulting time resolution is given by:
\begin{equation}
\Delta t \approx \frac{\delta t_{\mathrm{IN}}}{\sqrt{N_\mathrm{phe}}}~.
\end{equation}
The width $\delta t_{\mathrm{IN}}$ is about 1\,ns for $\gamma$-ray pulses, a few ns for hadron pulses, for muons a few hundred ps and about 2--4\,ns for the calibration pulses.
\item The transit time spread $\delta t_\mathrm{TTS}$ of the photo-multiplier (the spread of the times between the release of an electron from the photo cathode and the corresponding signal at the PMT output) which can be of the order of a few hundred ps per single photo-electron, depending on the wavelength of the incident light. As in the case of the photon arrival time spread, the total time spread scales with the inverse of the square root of the number of photo-electrons:
\begin{equation}
\Delta t \approx \frac{\delta t_{\mathrm{TTS}}}{\sqrt{N_\mathrm{phe}}}~.
\end{equation}
\item The reconstruction error due to the background noise and limited extractor resolution: 
%This contribution is inversely proportional to the signal to square root of background light intensities.
%
\begin{equation}
\Delta t \approx \frac{\delta t_{\mathrm{rec}} \cdot R/\mathrm{phe}}{N_\mathrm{phe}}
\end{equation}
where $R=\sqrt{\mathrm{Var}[\widehat{N}_\mathrm{phe}]}$ is the square root of the extractor variance, which depends only very weakly on the signal charge. %defined in equation~\ref{eq:def:r}

\item A constant offset due to the residual FADC clock jitter between different channels
%residual FADC clock jitter
 or the MC simulation time steps.
\begin{equation}
\Delta t \approx \delta t_0~.
\end{equation}
\end{enumerate}
In total, the time spread can be expressed as:
\begin{equation}
\Delta T = \sqrt{\frac{T_1^2}{N_\mathrm{phe}} + \frac{T_2^2}{N^2_\mathrm{phe}} + T_0^2}~.
\label{eq:Extraction:timefit}
\end{equation}
where $T_1$ contains the contributions of $\delta t_{\mathrm{IN}}$ and $\delta t_{\mathrm{TTS}}$, the parameter $T_2$ contains the contribution of $\delta t_{\mathrm{rec}}$ and $T_0$ the offset $\delta t_0$.

The measured time resolutions in figure~\ref{fig:Extraction:timedep} were fitted 
by equation (\ref{eq:Extraction:timefit}). 
%The results of the fit are summarized in table~\ref{tab:Extraction:timefitresults}. 
The low fit probabilities are partly due to the systematic differences in the intrinsic pulse shapes of the different color LED light pulses. Nevertheless, all calibration colors had to be included in the fit to cover the full intensity range. In general, the time resolutions for the UV pulses are systematically better than those for the other colors. This can be attributed to the fact that the UV pulses have a smaller intrinsic pulse width \cite{markus} and the UV LEDs are very stable from event to event, whereas the green and blue LED pulses can show secondary pulses about 10--40\,ns after the main pulse (see section \ref{sec:Extraction:nphe}), which influence the reconstructed pulse arrival time.

There are clear differences between the studied time extractors, especially the sliding window extractor yields poorer resolutions. %The spline and the digital filter are compatible within the errors. 
This is in part due to the fact that in the chosen sliding window algorithm the bias of the time reconstruction with respect to the relative timing (phase) of the pulses with respect to the free running FADC clock has not been corrected for.
The parameters $T_1$ and $T_0$ should in principle be independent of the time extraction 
algorithm. Nevertheless, $T_1$ is larger for the sliding window algorithm than for the 
spline interpolation and the digital filter. This is in part due to the (anti)-correlation 
between the reconstructed charge and arrival time for the former extractor, 
see equation~(\ref{eq:sliding_window_time}).

From the measured time resolution for calibration pulses one can estimate the expected time resolution for cosmic pulses. The only important difference between calibration and cosmic pulses are different arrival time spreads of the photons on the PMT camera. The time spread of the photons on the PMT for cosmic pulses is smaller than for blue/green calibration pulses, but about the same as for the UV calibration pulses (see the widths of the pulses in figure 3b). \footnote{Note, that the calibration light pulses illuminate directly the camera, whereas the cosmic light pulses are reflected by the MAGIC mirror system. The MAGIC mirrors have been built in a parabolic shape, and are thus isochronous. Nevertheless, they have been staggered in a chess-board manner \cite{Magic_mirrors} with an offset of about 10~cm. This introduces an additional contribution of about 700\:ps width to the intrinsic arrival times spread of the Cherenkov photons.} 
%The pulse sh cosmic pulse shown in figure 3b 
%In the case of cosmic pulses the arrival time spread of the photons on the PMT camera
% There are two differences between calibration and cosmics pulses: Different intrinsic arrival time spreads of the photons from calibration pulses and on the PMT
%these two cases, which influence the time resolution: The intrinsic arrival time spread of the photons on the PMT for cosmic pulses is about the same as for the UV calibration pulses, but smaller than for blue/green calibration pulses (see the widths of the pulses in figure 3b). 
%
%Contrary to cosmic pulses, the calibration light pulses illuminate directly the camera, they are not reflected in the MAGIC mirror system. In order to get an upper limit for the reconstructed arrival time resolution of cosmic pulses, the effect of the mirrors has to be included. The MAGIC mirrors have been built in a parabolic shape, and are thus isochronous. Nevertheless, they have been staggered in a chess-board manner \cite{Magic_mirrors} with on offset of about 10~cm. This introduces an additional %, binomial distributed contribution (about 700\:ps width) contribution about 700\:ps width to the intrinsic arrival times of the Cherenkov photons. Taking this effect into account, the expected time spread for inner pixels and cosmic pulses can be conservatively estimated to:
Therefore, the timing resolution for cosmic pulses is at least at the level of the timing resolution determined from the calibration pulses. The timing resolution for cosmic pulses is conservatively estimated to: 
\begin{equation}
\Delta T_{\mathrm{cosmic}} \approx \sqrt{\frac{4.5\,\mathrm{ns}^2}{N_\mathrm{phe}} 
+ \frac{20\,\mathrm{ns}^2}{N^2_\mathrm{phe}} + 0.04\,\mathrm{ns}^2} .
\label{eq:Extraction:timefitprediction}
\end{equation}
For signal charges above 10 photo-electrons the time resolution is below 830~ps. For signals of 100 photo-electrons the time resolution may be as good as 300~ps.

\section{CPU Requirements \label{sec:speed}}

The speed of different extractor algorithms (the number of reconstructed events per unit time) was measured on an Intel\ Pentium\ IV, 2.4\,GHz~CPU machine. Table~\ref{tab:Extraction:cpu} shows the average results whereby the individual measurements could easily differ by about 20\% from one try to another (using the same extractor). The numbers in this list have to be compared with the event reading and decompression speed (400~events/s). Every signal extractor being faster than this reference number does not limit the total event reconstruction speed. Only some of the integrating spline extractor configurations lie below this limit and would need to be optimized further.

\begin{table}[htp]
\centering
\begin{tabular}{|c|c|}
%\hline
%\hline
%\multicolumn{3}{|c|}{\large Measured Extraction Speed} \ruu\ruo \\
\hline
\hline
 Name         &  Events/s. \\
              &  (CPU)       \\
\hline                                                     
\hline                                                     
%Fixed Window 14 slices & 2700--3300  \ruu\ruo \\   
Fixed Window 8 slices & 3200--4000  \ruu\ruo \\   
\hline
% Sliding Window 2 slices & 400--700 & \ruo \\   
% Sliding Window 4 slices & 500--800 & \\   
Sliding Window 6 slices & 1000--1300  \\   
% Sliding Window 8 slices & 1100--1400 & \ruu \\   
\hline                          
Spline Amplitude      & 700--1000  \ruo \\
Spline Integral 1 sl. & 300--500  \\
% Spline Integral 2 sl. & 200--400 & \\
% Spline Integral 4 sl. & 150--200 & to be optimized \\
% Spline Integral 6 sl. & 80--120 & to be optimized \ruu \\
\hline                               
Digital Filter        & 700--900    \ruo \\
% Digital Filter 6 slices & 700--900 &   \ruo \\
% Digital Filter 4 slices  & 700--900 &  \ruu \\
\hline
\hline
\end{tabular}
\caption[Extraction speed different signal extractors]{%
\label{tab:Extraction:cpu}
The extraction speed measured for different signal extractor configurations. Note, that the fixed window does not calculate the arrival time.}
\end{table}

\section{Results and Discussion\label{sec:Extraction:results}}

The results based on the investigations discussed above are summarized in table~\ref{tab:Extraction:result}. Note, that there is no absolute basis for criteria to separate acceptable from non-acceptable properties of signal extraction algorithms. 
%Note, that although these criteria a well-motivated, there is no absolute basis to separate acceptable from non-acceptable properties
%The following criteria are used to compare the extractors:
In the following the arbitrarily chosen criteria to compare the extractors are motivated:

\begin{itemize}
\item The extractor should yield on average the true number of photo-electrons and should not deviate by more than 10\% in case of slight modifications of the pulse shape. These deviations directly effect the determination of the absolute energy scale of the reconstructed events. Note, that the dominant systematic error to the absolute energy scale is currently the photon detection efficiency (10-12\%) \cite{MAGIC_Crab}.
This requirement excludes extractors which integrate only a small portion of the pulse, especially the amplitude sensing cubic spline extractor. 
\item The extractor must yield a stable low gain pulse extraction. This means that apart 
from being robust against modifications of the pulse shape, the extractor has to 
reconstruct on average the true signal charge also in case of variations of the pulse 
position within the recorded FADC samples. This criterion excludes the fixed window 
extractor since arrival time jitters may exceed the time window between the tail of the 
high gain pulse and the beginning of the low gain pulse. 
%In certain data taking periods, also extractors using large sliding windows, are excluded by this criterion.
%% \item The extractor must also yield on average the correct charges for the low gain pulses on average.
%% \item The reconstructed charge must be linear to the input signal charge for all signals above the 
%% image cleaning level and below the low gain saturation level.
\item The \textit{RMSE}  of the reconstructed charge for the case of no signal should not exceed 2 photo-electrons (an arbitrarily chosen threshold) for dark night observations and the \textit{RMSE} of the reconstructed charge for air shower signals should never exceed the intrinsic Poissonian signal fluctuations plus excess noise above 5 photo-electrons. Camera pixels with a signal below 5 photo-electrons are usually rejected for the image parameterization \cite{Hillas_parameters,Fegan1997}. This low-energy analysis condition discards the large sliding windows and the fixed window extractor. It is not critical for high-energy analyses, however.
%% \item The variance of the reconstructed charge should not exceed twice the variance of the best extractor.
\item For analyses close to the energy threshold, an extractor should have a small or negligible bias, discarding again the amplitude sensing cubic spline extractor.
\item The time resolution should not be worse than 2\:ns at a signal strength of 10 photo-electrons. Note, that this condition allowed us to require a time coincidence of 3.3~ns between neighboring pixels to reject noise signals in the image cleaning and thus allow to measure differential energy spectra down to 60 GeV \cite{MAGIC_Crab}. All fixed window and all simple sliding window extractors are excluded by this condition.
%% \item The number of mis-reconstructed times should not exceed 0.5\% on average (including the FADC jumps).
\item The needed CPU-time should not exceed the one required for reading the data into memory and writing it to disk. Unless further effort is made to speed up the integrating spline, it is excluded if used with a large integration window.
\end{itemize}

\newcommand{\no}{\textcolor{red}{\bf NO\xspace}}
\newcommand{\ok}{\textcolor[rgb]{0.,.7,0.}{\bf OK\xspace}}
\newcommand{\best}{\textcolor[rgb]{0.,.4,0.}{\bf BEST\xspace}}

\begin{table}[htp]
\small{%
%\rotatebox{90}{%
\centering
\begin{tabular}{|l|c|c|c|c|c|c|}
%\hline
%\hline
%\multicolumn{8}{|c|}{\large Tested Signal Extractor Characteristics} \rule{0mm}{6mm} \rule[-2mm]{0mm}{6mm} \hspace{-3mm}\\
\hline
\hline
Extractor    & robust-  & robust-  & \textit{RMSE} & bias & time   & Speed \ruo \\
Configuration& ness     & ness     &      &      & spread &  \\
             & pulse       & pulse       &      &      &  & \\
             & form        & form        &      &      &        & \\
             & high gain   & low gain    &      &      &        & \ruu \\
\hline                                                     
\hline                                                     
%Fixed Window 14 sl.   & \best & \ok & \no  & \best & \no  & \best \ruo\ruu \\   
Fixed Window 8 sl.   &  \ok & \no & \no  & \best & \no  & \best \ruo\ruu \\   
\hline                                                         
Sliding Window 2 sl.  & \no & \no & \ok  & \ok & \no  & \ok \ruo \\   
Sliding Window 4 sl.  & \ok & \no & \no  & \ok & \no  & \ok \\   
Sliding Window 6 sl.  & \ok & \ok & \no  & \ok & \no  & \ok \\   
Sliding Window 8 sl.  & \ok & \best& \no  & \ok & \no  & \ok \ruu \\   
\hline                                                               
Spline Amplitude      & \no & \no & \ok  & \no & \ok   & \ok \ruo \\
Spline Integral 1 sl. & \no & \no & \ok  & \ok & \best & \ok \\
Spline Integral 2 sl. & \no & \no & \ok  & \ok & \best & \ok \\
Spline Integral 4 sl. & \ok & \no & \ok  & \ok & \best & \no \\
Spline Integral 6 sl. & \ok & \ok & \no  & \ok & \best & \no \ruu \\
\hline                                                                     
Digital Filter 4 sl.  & \ok & \no & \best& \ok & \ok  & \ok \ruo \\
Digital Filter 6 sl.  & \ok & \ok & \ok  & \ok & \ok & \ok \ruu \\
\hline
\hline
\end{tabular}
%}
\caption[Results signal extractor tests]{%
\label{tab:Extraction:result}
The tested characteristics for every extractor. See text for descriptions of the individual columns.
 \ok\ means, the extractor has passed the 
test, \no\ that the extractor failed and \best\ that the extractor has succeeded a particular 
test as best of all.}
}
\end{table}

Table~\ref{tab:Extraction:result} shows which extractors satisfy the above criteria. 
One can see that there is no signal extractor without problems. However, the digital 
filter fitting four FADC slices can always be used for the high gain extraction, and 
the digital filter fitting six FADC slices for the low gain extraction.%, whenever 
the mean pulse position is not critical. 
This combination has been chosen as the standard extractor for all MAGIC data before April 2007 with the 300 MSamples/s FADC system. 
During a certain period, the pulse position was by mistake shifted with respect to the FADC read-out samples. In this case \cite{MAGIC_1218} the signal was reconstructed by the cubic spline algorithms integrating 1--2 FADC slices.

% It has turned out to be robust, except for the data affected by the pulse position problem. 
%For the critical data, only the cubic spline algorithms integrating 1--2 FADC slices are left which might nevertheless yield systematic differences due to the instability with respect to deviations of the pulse form. In fact, data taken in certain periods of 2005 had to be extracted this way.

If efficiencies at low energies are not critical, i.e. a high analysis energy threshold without the use of the timing information, the sliding window extractor can be used in configurations which cover the entire pulse. This extractor turns out to be especially robust and was used for the data analysis in \cite{MAGIC_501}.% (whenever the mean pulse position is not critical). 

\section{Conclusions and Outlook} \label{sec:outlook}

In this paper different algorithms to reconstruct the charge and arrival time from the FADC read-out samples of the MAGIC telescope have been developed. These algorithms are tested using MC simulations, pedestal and calibration events. The achievable charge and arrival time resolutions are determined. A digital filter fitting four FADC slices in the high gain channel and six FADC slices in the low gain was chosen as the standard signal extraction algorithm.

Part of the difficulties to find a suitable signal extractor (reflected in 
table~\ref{tab:Extraction:result}) stem from the fact that the MAGIC signals are shaped 
just as long as to cover about four FADC slices. This choice, although necessary for a 
300~MSamples/s FADC read-out, ``washes out'' the intrinsic pulse form differences between 
$\gamma$-ray  and hadron  showers, and thus prevents the analysis from using this 
information in the $\gamma$/hadron discrimination, see \cite{Tescaro2005}. On the other side, the shaping time is 
not long enough to safely extract the amplitude from the (shaped) signal. 

These problems will be overcome with the full analysis of data taken with the 
new 2\:GHz FADC system in MAGIC \cite{GSamlesFADC,domino}; for first results see \cite{Tescaro2007}. This system has been designed 
to reduce any pulse form deformation to the minimum. It can be expected that the 
individual pulse forms are then directly recognized as such, e.g. with a digital filter 
using two sample pulse forms (a $\gamma$-ray-like and a hadron-like) and discriminating 
between the two with the help of the calculated $\chi^2$. These FADCs have a higher dynamic range and do not need a separate low gain channel any more. It can be expected that the signal extraction will become more robust, besides extracting a wealth of additional information about the shower characteristics.

%------------------------------------------------------------------------------

\appendix

\section{Acknowledgements}

We would
like to thank the IAC for the excellent working conditions at the
Observatory de los Muchachos in La Palma. The support of the
German BMBF and MPG, the Italian INFN and the Spanish CICYT is
gratefully acknowledged. The MAGIC telescope is also supported by ETH
Research Grant TH~34/04~3 and the Polish MNiI Grant 1P03D01028.

\end{document}